\documentclass[a4paper,fleqn]{cas-dc}

\usepackage[numbers, sort&compress]{natbib}

\def\tsc#1{\csdef{#1}{\textsc{\lowercase{#1}}\xspace}}
\tsc{WGM}
\tsc{QE}
\tsc{EP}
\tsc{PMS}
\tsc{BEC}
\tsc{DE}

\usepackage[english]{babel}
\usepackage{latexsym}
\usepackage{epsfig}

\usepackage{graphicx}
\usepackage{psfrag}
\usepackage{subfigure}
\usepackage{amsfonts}
\usepackage{amsmath, amssymb}
\usepackage{setspace}
\usepackage{pgfplots}
\usepackage{pgfplotstable}
\usepackage{tikz}
\usepackage{colortbl}
\usepackage{keyval}
\usepackage{xcolor}
\usepackage{enumerate}
\usepackage{algorithm}
\usepackage{algpseudocode}
\usepackage{xspace}
\usepackage{balance}
\usepackage{listings}
\usepackage{amssymb}
\usepackage{microtype}
\usepackage{caption}

\usetikzlibrary{patterns}

\newtheorem{theorem}{Theorem}
\newtheorem{lemma}{Lemma}

\newtheorem{constraint}{\textbf{Constraint}}

\newenvironment{proof}
{\vspace{12pt} \noindent {\bf Proof.} \\
	\noindent}{$\Box$ \vspace{12pt}}

\newcommand{\NEW}[1]{{\color{black} #1}}

\newcommand{\ceilfrac}[2]{\left\lceil\frac{#1}{#2}\right\rceil}

\newcommand{\introname}[1]{\textit{#1}}
\newcommand{\OFFR}[3]{O_{#1,#2}(#3)}
\newcommand{\EXECR}[3]{E_{#1,#2}(#3)}
\newcommand{\FINR}[3]{F_{#1,#2}(#3)}
\newcommand{\CR}[4]{C_{#1,#2}(#3,#4)}

\newcommand{\TASK}[1]{\tau_{#1}}

\newcommand{\COREC}[1]{p_{#1}}
\newcommand{\CANACCEL}[1]{\iota_{#1}}
\newcommand{\ISACCEL}[1]{a_{#1}}
\newcommand{\ACCELERATOR}{\mathcal{H}}

\newcommand{\PERIOD}[1]{T_{#1}}
\newcommand{\DEADLINE}[1]{D_{#1}}

\newcommand{\NONE}{\texttt{F}}

\newcommand{\SYNC}{\texttt{T}}

\newcommand{\TASKSET}{\Gamma}
\newcommand{\TASKSETK}[1]{\Gamma_{#1}}
\newcommand{\SSTASKSETK}[1]{\Gamma_{#1}^{\text{ss}}}

\newcommand{\WCETSHORT}[2]{C_{#1}^{#2}}
\newcommand{\SCHEDPOINTS}[1]{\mathcal{V}_{#1}}

\newcommand{\D}[1]{D_{#1}}

\newcommand{\PROCESSORSET}{{\mathcal{P}}}
\newcommand{\JITs}[2]{J_{#1}^{#2}}

\newcommand{\CHAIN}[1]{\omega_{#1}}
\newcommand{\CHAINSET}{\Omega}

\newcommand{\HP}[2]{hp_{#1}(#2)}
\newcommand{\HPP}[2]{hp^*_{#1}(#2)}
\newcommand{\HPSET}[2]{hp_{#2}(#1)}
\newcommand{\LPSET}[2]{lp_{#2}(#1)}
\newcommand{\EELATENCY}[1]{L_{#1}}

\newcommand{\WCETMIN}[1]{C^{\text{\footnotesize MIN}}_{#1}}

\newcommand{\WCETMINHW}[1]{C^{\text{\footnotesize MIN-HW}}_{#1}}

\newcommand{\SSTASK}[1]{\tau^{\text{ss}}_{#1}}

\newcommand{\R}[2]{R_{#1}^{#2}}

\newcommand{\WCETACC}[1]{C_{#1}^{\mathit{A}}}
\newcommand{\WCETNACC}[1]{C_{#1}^{\mathit{NA}}} 
\newcommand{\PROCVAR}[2]{\text{TC}_{#1,#2}} 
\newcommand{\SAMEPROC}[2]{\text{SC}_{#1,#2}} 
\newcommand{\SAMEPROCK}[3]{\text{SCC}_{#1,#2,#3}} 
\newcommand{\ACV}[1]{\text{AS}_{#1}} 
\newcommand{\INTERFWCET}[2]{\text{EI}_{#1,#2}} 
\newcommand{\SR}[1]{\text{STS}_{#1}} 
\newcommand{\ST}[1]{\text{ST}_{#1}} 
\newcommand{\LA}[1]{\text{LA}_{#1}} 
\newcommand{\PRIO}[2]{\text{TP}_{#1,#2}}
\newcommand{\PRIOABS}[1]{\text{PR}_{#1}}
\newcommand{\HIGHERPRIO}[2]{\text{HP}_{#1,#2}}

\newcommand{\ETASK}[1]{\text{ET}_{#1}} 
\newcommand{\ESEGM}[2]{\text{ES}_{#1,#2}}

\newcommand{\RTMAX}{\text{RT}_{\text{MAX}}}
\newcommand{\RTVAR}[1]{\text{RT}_{#1}}
\newcommand{\RTCVAR}[1]{\text{RTC}_{#1}}
\newcommand{\ENVAR}[1]{\text{SV}_{#1}}

\newcommand{\EACCVAR}[2]{\text{EH}_{#1, #2}}
\newcommand{\EACCRVAR}[3]{\text{EHR}_{#1, #2, #3}}
\newcommand{\STCVAR}[1]{\text{STC}_{#1}}
\newcommand{\ENSUSPVAR}[1]{\text{SS}_{#1}}
\newcommand{\BLOCKVAR}[1]{\text{BL}_{#1}}

\newcommand{\LAT}[1]{\text{L}_{#1}}
\newcommand{\MAXLAT}{\text{L}_{\text{MAX}}}

\newcommand{\REGION}[2]{\rho_{#1,#2}}
\newcommand{\REGIONSET}[1]{\rho_{#1}}
\newcommand{\ACCREGIONSET}[1]{\rho_{#1}^H}

\newcommand{\BIGM}{M}

\newcommand{\RELPATTERN}{\Lambda}

\begin{document}
\let\WriteBookmarks\relax
\def\floatpagepagefraction{1}
\def\textpagefraction{.001}
\shorttitle{{\footnotesize Optimized Partitioning and Priority Assignment of Real-Time Applications on  Heterogeneous Platforms with Hardware Acceleration}}
\shortauthors{Daniel Casini et~al.}

\title [mode = title]{Optimized Partitioning and Priority Assignment of Real-Time Applications on  Heterogeneous Platforms with Hardware Acceleration}

\author[1,2]{Daniel Casini}[orcid=0000-0003-4719-3631]\cormark[1]\ead{daniel.casini@santannapisa.it}

\author[3]{Paolo Pazzaglia}[orcid=0000-0003-0377-3327]\ead{pazzaglia@cs.uni-saarland.de}

\author[1,2]{Alessandro Biondi}[orcid=0000-0002-6625-9336]\ead{alessandro.biondi@santannapisa.it}

\author[1,2]{Marco Di Natale}[orcid=0000-0002-4480-8808]\ead{marco.dinatale@santannapisa.it}

\address[1]{TeCIP Institute, Scuola Superiore Sant'Anna, Via G. Moruzzi 1, 56124 Pisa (PI), Italy  }
\address[2]{Department of Excellence in Robotics \& AI, Scuola Superiore Sant'Anna, Via G. Moruzzi 1, 56124 Pisa (PI), Italy  }
\address[3]{Saarland University, Saarland Informatics Campus, E1.3, 66123 Saarbr\"ucken, Germany}

\cortext[cor1]{Corresponding author}

\begin{abstract}  
Hardware accelerators, such as those based on GPUs and FPGAs, offer an excellent opportunity 
to efficiently parallelize functionalities. 
Recently, modern embedded platforms started being equipped with such accelerators, resulting in a compelling choice for emerging, highly computational intensive workloads, like those required by next-generation autonomous driving systems. 
Alongside the need for computational efficiency, such workloads are commonly characterized by real-time requirements, which need to be satisfied to guarantee the safe and correct behavior of the system. 
To this end, this paper proposes a holistic framework to help designers partition real-time applications on heterogeneous platforms with hardware accelerators. The proposed model is inspired by a realistic setup of an advanced driving assistance system presented in the WATERS 2019 Challenge by Bosch, further generalized to encompass a broader range of  heterogeneous architectures.
The resulting analysis is linearized and used to encode an optimization problem that jointly \textbf{(i)} guarantees timing constraints, \textbf{(ii)} finds a suitable task-to-core mapping, \textbf{(iii)} assigns a priority to each task, and \textbf{(iv)} selects which computations to accelerate, seeking for the most convenient trade-off between the smaller worst-case execution time provided by accelerators and synchronization and queuing delays.
\end{abstract}


\maketitle

\section{Introduction}

Embedded real-time systems have been subject to considerable changes over the last two decades. First, the advent of multi-core platforms introduced new allocation and scheduling problems~\cite{DavisSurvey} and the consideration of contention delays on shared resources such as memories~\cite{GracioliSurvey, MaizaSurvey} and I/O devices~\cite{Casini2021RTAS}.
More recently, the race towards feature-rich, predictable, safe, and secure autonomous vehicles shifted the attention to devices capable of performing a huge amount of parallel computation in an efficient way: \emph{heterogeneous platforms}.
 
Heterogeneous platforms are composed of multiple cores, possibly with different characteristics. Often, they are also provided with \emph{hardware accelerators}, such as graphic processing units (GPUs), field-programmable gate arrays (FPGAs),  or digital signal processors (DSPs). 
The accelerators have proven to be an essential means to feasibly implement the perception and prediction software required by autonomous cars~\cite{Bateni2020}. 
Indeed, such functionalities commonly require the usage of deep neural networks and computer vision algorithms that cannot be efficiently executed by processor cores. 
 
However, improving the timing efficiency by means of hardware accelerators is just a piece of the puzzle. 
The new software introduced for autonomous driving is subject to real-time constraints, and it is required to interact and communicate \emph{in a predictable way} with all the other time critical (and often legacy) software for the control of the vehicle~\cite{Wurst2019}.
This aspect calls for action on the side of the real-time scheduling analysis.
For example, a computation may start on a processor core, continue on a hardware accelerator, and complete again on a core. 
Clearly, these behaviors require a richer modeling and more complex analysis strategies.

Furthermore, the engineers are left with several design choices, e.g., deciding the best task-to-core mapping, and deciding whether to use hardware accelerators when multiple implementations of the same functionality are available. 
This process is not trivial, and becomes even more complicated when it is required to guarantee that the timing constraints on the whole application are respected. 
For example, is it best to execute a computation slowly on a fairly empty CPU, or faster in a congested GPU? Such choices are critical and can heavily affect the performance of the system. 

A possible approach to solve this issue may involve experimenting with different configurations, and empirically observing the results, seeking the best performing one. However, this exhibits the following shortcomings: \textbf{(i)} it may be highly time-consuming, \textbf{(ii)} it only allows to check a small number of configurations, without reaching holistic conclusions, \textbf{(iii)} it provides no real-time guarantees and predictability, and \textbf{(iv)} it is very unlikely to be optimal or near-optimal. 

These drawbacks give rise to the need for off-line design and analysis strategies, to guarantee that all the tasks fulfill their timing constraints while taking the best decision on multiple design choices, e.g., whether to accelerate a task or not, how to allocate tasks to cores, and how to assign priorities.

While pursuing this goal, one may be tempted to rely on overly simplistic models, leading to analysis and design strategies useful from a theoretical point of view but possibly far from reality or, on the opposite side, to solve a specific problem on a specific computing platform, achieving a result that does not generalize to other cases, thus limiting its usefulness for the research community.

\smallskip
\noindent \textbf{This Paper.} To avoid falling into the two aforementioned issues, this paper tries to balance between generality and specificity. To this end, we start by looking at a specific and realistic problem, the WATERS 2019 Challenge proposed by Bosch~\cite{WATERS2019}, which provides data (execution times, communication relations, etc.) for an Advanced Driver-Assistance System (ADAS) application running on a \textit{NVIDIA Jetson TX-2}, a heterogeneous platform  with a GPU accelerator. Based on this system configuration, we build a model to analyze its real-time behavior while generalizing to other processing platforms with other kinds of accelerators. Then, we show how to build optimization strategies to determine suitable acceleration decisions, task allocations, and priority assignments, which are then evaluated on the WATERS Challenge model in our experimentation. 

This paper builds upon the workshop paper from the same authors~\cite{casini2019addressing} presented at the 10th International Workshop on Analysis Tools and Methodologies for Embedded and Real-time Systems (WATERS 2019), as a solution to the WATERS 2019 Challenge.
The original contribution~\cite{casini2019addressing} is extended in several directions: \textbf{(i)} the framework used in this paper now supports tasks performing acceleration in multiple separate segments of their execution; \textbf{(ii)} we study two different scheduling policies for the accelerator, i.e., round-robin and non-preemptive fixed priority, and provide the corresponding analyses; \textbf{(iii)} we propose a response time analysis for self-suspending tasks that is suitable for linear optimization, inspired by (and extending) the approach presented in~\cite{Pazzaglia2019DATE}; \textbf{(iv)} we provide a comprehensive optimization problem, extensively discussing all the constraints and presenting a corresponding proof for each of them; \textbf{(v)} we provide a comprehensive evaluation based on the WATERS Challenge model that studies different objective functions and scheduling policies. 

\smallskip
\noindent \textbf{Paper Structure.} The remainder of the paper is organized as follows. Section~\ref{sec:waters} introduces the WATERS 2019 Challenge. Section~\ref{sec:model} presents our modeling solution to describe an application running on a heterogeneous platform with a hardware accelerator. Section~\ref{sec:analysis} presents methods to analyze heterogeneous applications using self-suspending task theory. Section~\ref{sec:partitioning} illustrates an optimization problem to take several design-level decisions in an optimal way. Section~\ref{sec:evaluation} presents the results of our experimental evaluation.
Section~\ref{sec:related} discusses the related work.  Finally, Section~\ref{sec:conclusions} concludes the paper.
\section{The WATERS 2019 Challenge}
\label{sec:waters}
 
\begin{figure}[t]	
	\centering	
	\includegraphics[width = 0.9\columnwidth]{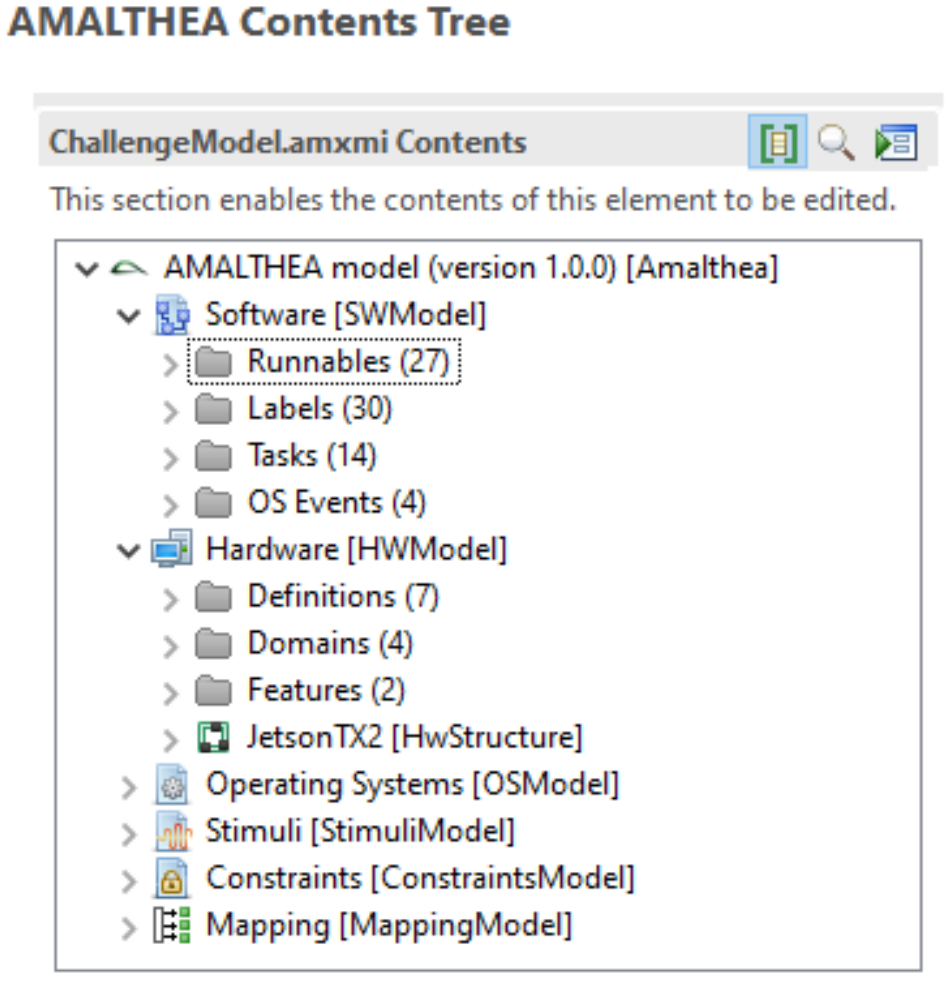}	
	\caption{Amalthea model provided with the WATERS 2019 Challenge. }
	\label{fig:amalthea}	
\end{figure}

\begin{figure*}
	
	\centering
	
	\includegraphics[scale=0.8]{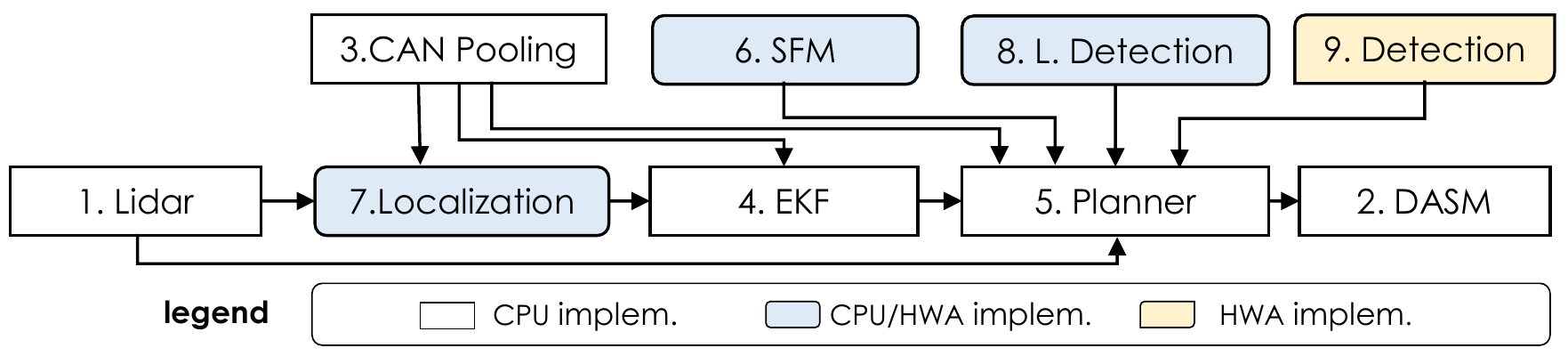}
	
	\caption{Processing chains of the WATERS 2019 Challenge. The number on the side of each task  represent the task ID, imported from the challenge data. }\label{fig:chains}
	
\end{figure*}

The WATERS 2019 Challenge~\cite{WATERS2019} by Bosch represents an interesting opportunity to explore a realistic design of a modern ADAS application, implemented on a heterogeneous platform. 
The Challenge provides an Amalthea APP4MC model (see Figure~\ref{fig:amalthea}) representing an ADAS prototype.
The application is composed of nine tasks performing computations from the sensors input to the steering command. 
Tasks have communication dependencies as shown in Figure~\ref{fig:chains}.
The model uses the Jetson TX-2 as a reference platform, with six cores organized in two processor islands.
The first island includes four ARMv8 A57 cores running at 1.9 GHz, while the latter contains two 2 Ghz ARMv8 Denver cores.  
The platform is also provided with an iGPU (integrated GPU), which allows accelerating some strongly-parallel computations. 

The Amalthea APP4MC model includes additional information about the structure of the tasks. For each task, periods and deadlines are specified.
Each task is then composed of multiple segments of computation, called \emph{runnables}, according to the AUTOSAR standard~\footnote{The AUTOSAR standard, version 4.3. \url{http://www.autosar.org}}. Each runnable implements a specific function and is characterized by a set of possible execution time values, depending on where it is implemented. If the runnable is designed to run on a CPU, it presents the execution time information for the A57 cores and for the Denver cores; if it is designed to be accelerated, it includes the execution time information computed for the GPU. 
In both cases, the minimum, average, and maximum execution times are reported. 
The model also specifies which \emph{labels} (i.e., shared memory locations) a runnable is reading or writing.  
This information is particularly useful for deriving the communication dependencies.
  
For each task that can be accelerated, the Amalthea model provides two task objects. 
The first one contains the preprocessing and postprocessing runnables that are executed on a CPU when the main activity is executed on the GPU: such additional runnables are required for passing the inputs to the GPU function and getting back the results.  The second object involves the runnables executing on the GPU and performing the computation. 
The WATERS 2019 challenge also provides (in another model file) an alternative version of some tasks, with their parameters in the case in which they are executed on a CPU (when allowed).   

Building upon this realistic case study, we derive a more general model that is used to analyze and optimize a generic real-time system running on a heterogeneous platform.  
The NVIDIA GPU device considered in the WATERS 2019 Challenge involves scheduling policies that are not publicly disclosed by the hardware vendor~\cite{Amert2017}, for which details may only be experimentally inferred through reverse engineering~\cite{Amert2017, Sanudo2020} or approximated~\cite{WATERS2019}. Conversely, we consider two predictable scheduling policies, round-robin and non-preemptive fixed-priority, which may be either directly adopted in the hardware accelerator or be enforced by an application-level scheduler that handles acceleration requests on the CPU side~\cite{Bio-RTSS16, Elliot2013}.

\section{System Model}
\label{sec:model}

The system model analyzed in the following of the paper builds upon the proposal the WATERS 2019 Challenge, and it is extended to make it representative of processing platforms with arbitrary heterogeneous cores and hardware accelerators.

Table~\ref{table:symbols} summarizes the main symbols used in this paper.

\subsection{Platform Model}  
This paper considers a heterogeneous embedded real-time platform composed of a set $\PROCESSORSET  = \{\COREC{1}, \ldots, \COREC{m}\}$ of processor cores.
Each core $\COREC{k} \in \PROCESSORSET$ is assigned a \emph{type} that determines the execution-time profile of the functionality implemented in that core. 
The type can be easily extended to cover other aspects of interest, such as power consumption and cache memory size, which however are out of the scope of this work.

For the sake of simplicity, we consider that the platform provides a single hardware accelerator, which is referred to as $\ACCELERATOR$. 
Nonetheless, the present analysis can be easily extended for the case of multiple independent accelerators at the expense of a more complex notation.

\subsection{Task Model} 

The application implemented in the platform comprises a set $\TASKSET = \{\TASK{1}, \ldots, \TASK{n}\}$ of periodic real-time tasks. Tasks are executed on cores according to a \introname{partitioned fixed-priority preemptive scheduling}, where each task $\TASK{i}$ is statically assigned to a processor $\COREC{k}$ and a unique priority $\pi_i$. This configuration guarantees a high predictability while being  representative of systems capable to run on hardware accelerators, i.e., those based on Linux, where partitioned fixed-priority scheduling can be achieved by assigning tasks to the \textsc{SCHED\_FIFO} scheduling class and specifying affinities with the \texttt{sched\_setaffinity()} system call.

We denote with $\TASKSETK{k} \subseteq \TASKSET$ the subset of tasks mapped on core $\COREC{k}$, with $\bigcup_k \TASKSETK{k} = \TASKSET$ and $\bigcap_k \TASKSETK{k} = \emptyset$.
 The set of all tasks with priority higher (resp. lower) than $\pi_i$ is denoted with $\HPSET{\TASK{i}}{}$ (resp. $\LPSET{\TASK{i}}{}$).   
Similarly, the set of all tasks mapped on core $\COREC{k}$ and with priority higher (resp., lower) than $\pi_i$ is denoted with $\HPSET{\TASK{i}}{k}$ (resp., $\LPSET{\TASK{i}}{k}$).
Each task $\TASK{i}$ releases a potentially infinite sequence of instances called \emph{jobs}, each separated by $\PERIOD{i}$ time units. Each job needs to complete within its relative deadline $\D{i}\leq\PERIOD{i}$, i.e., within $\D{i}$ units of time from its release.  
A task is said to be \emph{schedulable} if all of its jobs always complete within $\D{i}$ time units from its release.

Furthermore, each task $\TASK{i} \in \TASKSET$ is composed of a sequence of code \emph{segments} executed sequentially, with $ \REGIONSET{i}= \{\REGION{i}{1}, \ldots, \REGION{i}{w}\}$ denoting the set of all segments $\REGION{i}{j} \in \REGIONSET{i}$. 
A job of task $\TASK{i}$ starts with the execution of segment $\REGION{i}{1}$, and any other segment $\REGION{i}{j}$ with $1 < j \leq w$ starts executing only after the completion of $\REGION{i}{j-1}$. 
Each segment represents a functionally distinct fragment of code, and its implementation can be either provided for execution on processor cores, on the hardware accelerator, or both. 
To this end, each segment is characterized by an implementation type $\CANACCEL{i,j} \in \{\mbox{\texttt{CPU}, \texttt{HWA}, \texttt{CPU-HWA}}\}$, where $\CANACCEL{i,j} = \mbox{\texttt{CPU}}$ if the segment can only be executed on a core; $\CANACCEL{i,j} = \mbox{\texttt{HWA}}$ if the segment can only be executed on the hardware accelerator $\ACCELERATOR$, and $\CANACCEL{i,j} = \mbox{\texttt{CPU-HWA}}$ if two implementations are provided and hence the segment can be executed either on a core or on the accelerator. 
 
When both implementations on cores and the hardware accelerator are provided, it necessary to determine where the segment actually executes.  Hence, each segment is further characterized by a parameter $\ISACCEL{i,j} \in \{\SYNC,$ $ \NONE \}$ denoting if $\TASK{i}$ offloads its computations to the accelerator ($ \ISACCEL{i,j} = \SYNC$) or not ($ \ISACCEL{i,j} = \NONE$). Clearly, if $\CANACCEL{i,j} = \mbox{\texttt{CPU}}$, then  $\ISACCEL{i,j} = \NONE$, and if $\CANACCEL{i,j} = \mbox{\texttt{HWA}}$, $\ISACCEL{i,j} = \SYNC$.  
The set of accelerated segments of task $\TASK{i}$ is denoted with $\REGIONSET{i}^A = \{ \REGION{i}{j} ~|~ \ISACCEL{i,j} = \SYNC \}$), while the set of segments that \emph{may be accelerated} is denoted with $\ACCREGIONSET{i} = \{ \REGION{i}{j} ~|~ \CANACCEL{i,j} \in \{\mbox{\texttt{CPU-HWA}}, \mbox{\texttt{HWA}}\}\}$.

\subsection{Offloading Mechanism} 
We consider a \emph{synchronous}, \emph{suspension-based} offloading mechanisms to the hardware accelerator. Namely, when a segment $\REGION{i}{j}$ of $\TASK{i}\in \TASKSETK{k}$ is accelerated (i.e., $ \ISACCEL{i,j} = \SYNC$), it first executes the \emph{offloading phase}, i.e., it executes a first chunk of code on its processor $\COREC{k}$ to perform the initial operations and to prepare the data to offload to the accelerator.
When it completes, the task suspends on $\COREC{k}$, and its execution continues on $\ACCELERATOR$ in the \emph{processing phase}. Upon completion, the task is awakened on $\COREC{k}$, and retrieves the outputs produced by the accelerator, possibly executing further processing in the \emph{finalization phase}, which terminates the execution of $\REGION{i}{j}$.
This behavior is representative of most hardware accelerators, e.g., those based on GPUs~\cite{WATERS2019} and FPGAs~\cite{Bio-RTSS16}.
Conversely, when a segment is not accelerated, i.e., $\ISACCEL{i,j} = \NONE$, it performs only the processing phase on $\COREC{k}$.

For such three phases of an arbitrary segment $\REGION{i}{j}$, we introduce $\OFFR{i}{j}{\COREC{k}}$, $\EXECR{i}{j}{\COREC{k}}$, and $\FINR{i}{j}{\COREC{k}}$ to denote the worst-case execution times (WCETs) of the offloading, processing, and finalization phase, respectively. Note that, due to the platform heterogeneity, the WCETs depend on the type of the core $\COREC{k}$. The offloading and finalization phases of $\REGION{i}{j}$ have a positive WCET only when $\ISACCEL{i,j} = \SYNC$. In this case,  the offloading and finalization phase run on the core $\COREC{k}$ where the corresponding task is allocated, with WCETs $\OFFR{i}{j}{\COREC{k}}$ and $\FINR{i}{j}{\COREC{k}}$, respectively.  The execution phase instead runs on the hardware accelerator $\ACCELERATOR$, and its WCET is denoted by $\EXECR{i}{j}{\ACCELERATOR}$. When $\ISACCEL{i,j} = \NONE$, the segment is not accelerated and composed of the execution phase only, which occurs on its core $\COREC{k}$. 

\begin{figure}[t]
	\includegraphics[width = \columnwidth]{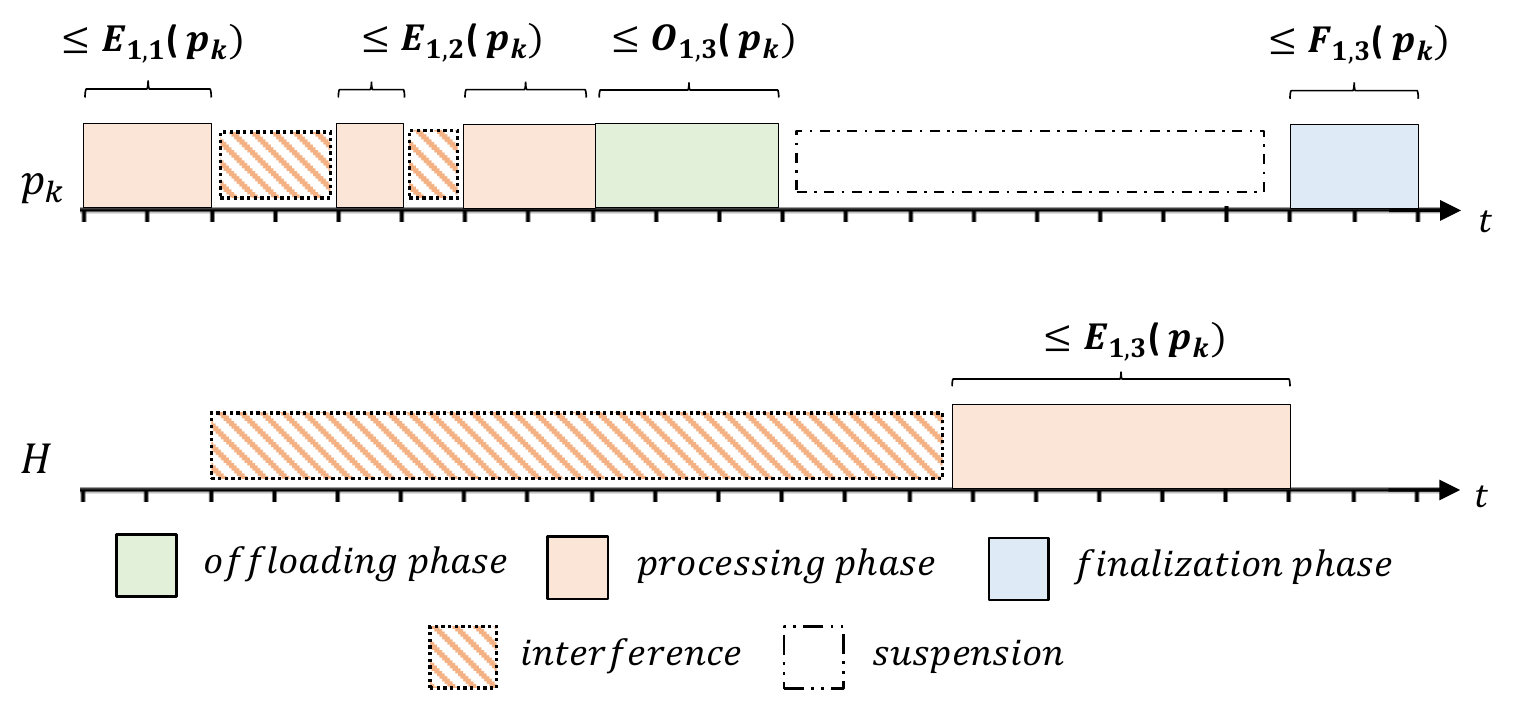}
	\caption{Example of schedule under the proposed model.}
	\label{fig:schedule}
\end{figure}

Figure~\ref{fig:schedule} shows an example of a possible schedule under the execution model of this paper. Task $\TASK{1}$ is composed of three segments. The first two are not accelerated, and hence they are only composed of a single processing phase each, executing on a core $\COREC{k}$, with durations bounded by the parameters $\EXECR{1}{1}{\COREC{k}}$ and $\EXECR{1}{2}{\COREC{k}}$, respectively. The third segment is accelerated. Therefore, it is composed of an offloading and finalization phase running on $\COREC{k}$ for at most $\OFFR{1}{3}{\COREC{k}}$ and $\FINR{1}{3}{\COREC{k}}$, respectively, and a processing phase running on the accelerator for at most $\EXECR{1}{3}{\ACCELERATOR}$. 

For brevity, the worst-case execution time of a segment $\REGION{i}{j}$ on a core $\COREC{k}$ is denoted as follows:
$$\CR{i}{j}{\COREC{k}}{\ISACCEL{i,j}} = \left \{ 
\begin{array}{ll}
\EXECR{i}{j}{\COREC{k}} & \mbox{if } \ISACCEL{i,j} = \NONE \\
\OFFR{i}{j}{\COREC{k}} + \FINR{i}{j}{\COREC{k}} & \mbox{if } \ISACCEL{i,j} = \SYNC
\end{array}
\right.$$ 
and the WCET of the whole task $\TASK{i}$, running on a core $\COREC{k}$, is hereafter referred to in compact notation as $\WCETSHORT{i}{}$, defined as follows:
$$\WCETSHORT{i}{} = \sum_{\REGION{i}{j}\in\REGIONSET{i}} \CR{i}{j}{\COREC{k}}{\ISACCEL{i,j}}.$$

Each task $\TASK{i}$ is also characterized by a worst-case response time (WCRT), which is the longest time span elapsed between the release and the completion of any of its jobs. Usually, the exact WCRT is difficult to derive, but an upper-bound $\R{i}{}$ can be found with a suitable response-time analysis. Then, if  $\R{i}{} \le \DEADLINE{i}$ for each task $\TASK{i}\in\TASKSET$, the system is said to be schedulable.
Methods to bound the WCRT under the configuration proposed in this paper are reported in Section~\ref{sec:analysis}.

\subsection{Task Chains} 

As discussed in Section~\ref{sec:waters}, tasks may have functional dependencies, e.g., producer-consumer relationships. This is modeled by denoting a sequence of communicating tasks with a \emph{processing chain} $\CHAIN{x}$, where $\CHAIN{x}$ represents an ordered list of tasks in a two-by-two producer-consumer relationship, i.e., $\CHAIN{x} = (\TASK{x_1}, \TASK{x_2}, \dots)$. As shown in Figure~\ref{fig:chains}, each task may belong to multiple chains, thus forming a graph of dependencies. The set of all the chains is denoted as $\CHAINSET$. In this paper, task chains are assumed to be time-triggered, i.e., each task $\TASK{i} \in \CHAIN{x}$ in the chain is periodically released according to its period $\PERIOD{i}$. 

In the next section, we discuss how to compute the end-to-end latency of such chains, and we introduce a response-time analysis for the tasks running on both the CPUs and the accelerator.

\begin{table}[h]
	\centering
	\small
	\caption{Table of symbols} 
	\label{table:symbols}
	\begin{tabular}{l|l}
		\hline
		\textbf{Symbol} & \textbf{Description} \\ \hline \hline
		$\COREC{k}$ & $k$-th processor core \\
		$\ACCELERATOR$ & hardware accelerator \\
		\hline
		$\TASK{i}$ & $i$-th task \\
		$\TASKSETK{k}$ & tasks assigned to $\COREC{k}$\\
		$\PERIOD{i}$ & $i$-th task period \\
		$\DEADLINE{i}$ & $i$-th task deadline \\
		$\R{i}{}$ & WCRT bound of $\TASK{i}$ \\
		\hline
		$\REGION{i}{j}$ & $j$-th segment of $\TASK{i}$ \\
		$\REGIONSET{i}$ & set of $\TASK{i}$'s segments \\
		$\CANACCEL{i,j}$ &  implementation type of $\REGION{i}{j}$ \\
		$\ISACCEL{i,j}$& equal to \SYNC~iff $\REGION{i}{j}$ executes on $\ACCELERATOR$ \\
		$\REGIONSET{i}^A$ & set of accelerated segments \\
		$\ACCREGIONSET{i}$ & segments that \emph{may} be accelerated \\
		\hline
		$\OFFR{i}{j}{\COREC{k}}$ & WCET of the offloading phase of $\REGION{i}{j}$ \\
		$\EXECR{i}{j}{\COREC{k}}$ & WCET of the processing phase of $\REGION{i}{j}$ \\
		$\FINR{i}{j}{\COREC{k}}$ & WCET of the finalization phase of $\REGION{i}{j}$ \\
		$\CR{i}{j}{\COREC{k}}{\ISACCEL{i,j}}$ & WCET of $\REGION{i}{j}$ on a core $\COREC{k}$ \\
		\hline
		$\CHAIN{x}$ & $x$-th chain \\
		$\CHAINSET$ & set of all chains \\
		\hline
	\end{tabular}
\end{table}

\section{End-to-End Latency Analysis}
\label{sec:analysis}

This section shows how to bound the end-to-end latency of task chains running on top of a heterogeneous platform, where each task is statically mapped in one of the CPU cores, but some of its computation can be offloaded to a hardware accelerator multiple times during execution.

We recall from prior work~\cite{Davare2007} that the end-to-end latency $\EELATENCY{x}$ of a (time-triggered) processing chain $\CHAIN{x}$ is bounded by: 
\begin{equation} \label{eqn:chainlate}
\EELATENCY{x} = \sum_{\TASK{i} \in \CHAIN{x}} (\R{i}{} + \PERIOD{i}) - \PERIOD{\mathit{\text{first}}},
\end{equation}
considering that the external event triggering the chain arrives synchronously with the release of the first task $\TASK{\text{first}}$ of the chain.
To apply Equation~\eqref{eqn:chainlate}, the worst-case response-time $\R{i}{}$ of each task in the chain needs to be bounded. 
Next, we show how to leverage existing results on \emph{self}-\emph{suspending tasks} theory~\cite{chen2019many} to analyze the behavior of a task performing acceleration on a core $\COREC{k} \in\PROCESSORSET$. 

\subsection{Self-Suspending Tasks}

To enable the presentation of the adopted analysis techniques, the segmented self-suspending task model is here briefly introduced. 

A segmented self-suspending task -- hereafter denoted with the symbol $\SSTASK{i}$ to better differentiate from the task model introduced in Section~\ref{sec:model} -- is characterized by an ordered sequence of $N^S_i$ \emph{regions} $(l_{i,1}, \ldots, l_{i,j}, \ldots, l_{i,N^S_i})$, representing alternating code executions and self-suspensions. 
Here we intentionally use the new concept of regions instead of segments, since they serve a different purpose, and will be used later to map the original task model of the paper to the current self-suspended tasks.
Both execution and suspension regions are characterized by a bounded worst-case duration.
If $l_{i,j}$ is an execution region, its WCET is denoted by $C_{i,j}$; otherwise, if $l_{i,j}$ is a suspension region, its duration is bounded by $S_{i,j}$.
Overall, the duration bounds of execution and suspension regions is  represented by the tuple 
$\langle C_{i,1}, S_{i,1}, \ldots, S_{i,N_i^S-1}, C_{i, N_i^S} \rangle$.
Analogously to Section~\ref{sec:model}, a self-suspending task
$\SSTASK{i}$ is periodically released with period $T_i$. Furthermore, each of them is characterized by a relative deadline
$D_i \leq T_i$ and fixed priority $\pi_i$.
Once a self-suspending task is released, the first execution region is also released.
If the $(j-1)$-th execution region of $\SSTASK{i}$ completes at time $t$, the $(j+1)$-th execution region 
is released no later than time $t+S_{i,j}$.
It is worth noting that a task $\SSTASK{i}$ may have no suspension regions but still be modeled as a self-suspending task: in that case, it will only consist of one execution region $l_{i,1}$ with execution time $C_{i,1}$.

\subsubsection{Mapping Accelerations to Self-Suspensions} 
\label{sssec:mapping} 
$\SSTASKSETK{k}$ indicates  the set of self-suspending tasks running on core $p_k$.
A task $\TASK{i} \in \TASKSETK{k}$ making use of hardware acceleration can be analyzed as a corresponding self-suspending task $\SSTASK{i} \in \SSTASKSETK{k}$ by establishing the following mapping. 

Given an arbitrary task $\TASK{i}\in\TASKSETK{k}$, each non-accelerated segment $\REGION{i}{j}$ (i.e., with $\ISACCEL{i,j} = \NONE$) is firstly mapped to an execution region of $\SSTASK{i} \in \SSTASKSETK{k}$ with WCET equal to $\EXECR{i}{j}{\COREC{k}}$. Conversely, an accelerated segment $\REGION{i}{j}$ ($\ISACCEL{i,j} = \SYNC$) is mapped as follows: 
\begin{enumerate}
\item the offloading phase of $\REGION{i}{j}$ is mapped to an execution region of $\SSTASK{i} \in \SSTASKSETK{k}$ with WCET equal to $\OFFR{i}{j}{\COREC{k}}$;
\item the processing phase is mapped to a suspension region of $\SSTASK{i} \in \SSTASKSETK{k}$;
\item the finalization phase is mapped to an execution region of $\SSTASK{i} \in \SSTASKSETK{k}$ with WCET equal to $\FINR{i}{j}{\COREC{k}}$.
\end{enumerate}
Finally, consecutive execution regions (i.e., not separated by a suspension region) are merged into a single region, and their WCETs are summed (as they all sequentially execute on the same core without suspensions). 
We denote by $\mathcal{Q}(\REGION{i}{j}) = l_{i,x}$ the suspension region that corresponds to segment $\REGION{i}{j}$, if any.

While the WCET of each execution segment of $\SSTASK{i}$ is thus known from the task parameters of $\TASK{i}$, the duration of each suspension region depends on the response time of the corresponding task executing on the hardware accelerator, considering that other segments may be contending the same computational resource. 
Consequently, bounding the maximum duration of a self-suspension regions involves bounding the response-time of each accelerated segment in the hardware accelerator.
Clearly, this requires knowing the scheduling policy adopted on the accelerator.  

\begin{figure}[t]
	\includegraphics[width = \columnwidth]{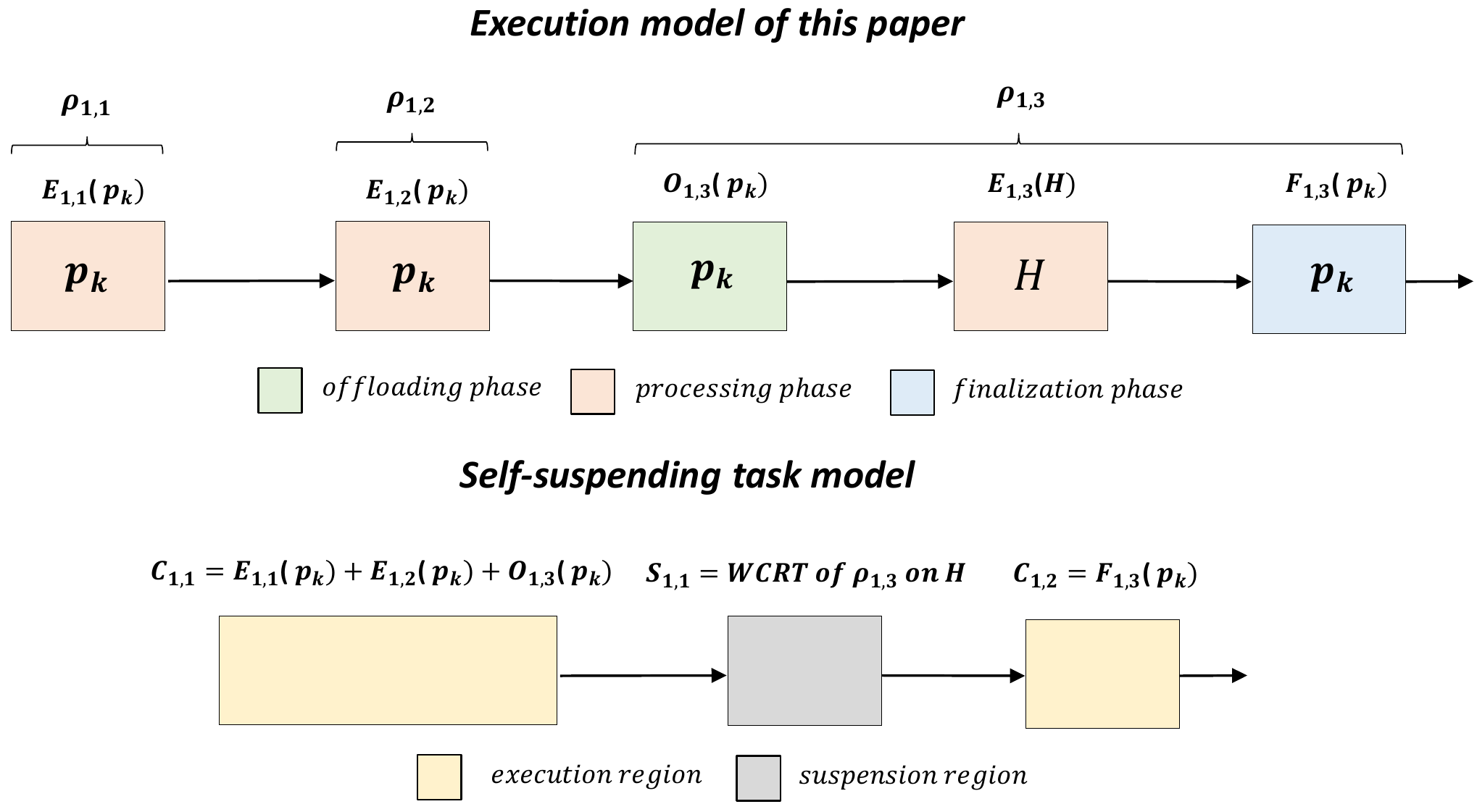}
	\caption{Example mapping of the model proposed in this paper to the self-suspending task model for analysis purpose.}
	\label{fig:Model}
\end{figure}

Figure~\ref{fig:Model} shows an example of mapping between the model proposed in this paper and the self-suspending task model. Task $\TASK{1}$ in the example is composed of three segments, $\REGION{1}{1}$, $\REGION{1}{2}$, and $\REGION{1}{3}$. Only the third one is accelerated. Hence, the corresponding self-suspending task $\SSTASK{1}$ running on $\COREC{k}$ is composed of an execution region with WCET $C_{i,1} = \EXECR{1}{1}{\COREC{k}} + \EXECR{1}{2}{\COREC{k}} +  \OFFR{1}{3}{\COREC{k}}$, a suspension region with length bounded by $S_{i,1}$, and an execution region with WCET $\FINR{1}{3}{\COREC{k}}$. The parameter $S_{i,1}$ is unknown beforehand and needs to be bounded by analyzing the worst-case response time of the accelerated segment running on $\ACCELERATOR$.

Before proceeding with the details of the accelerator scheduling, we discuss the response-time analysis for self-suspending tasks.

\subsection{Response-Time Analysis with Self Suspensions}
 
Several analysis techniques for self-suspending tasks are available in the literature (please refer to the work by Chen et al.~\cite{chen2019many} for a detailed review). Next, we explore a method for analyzing a self-suspending task which is particularly suited to be applied to our optimal mapping problem while still providing good schedulability performance. 

\NEW{In the following paragraph the analysis is performed considering a task set of self-suspending tasks $\SSTASKSETK{}$ obtained using the mapping of Section~\ref{sssec:mapping}  from the task set under analysis.}

\subsubsection{Jitter-based Analysis} \label{s:jitter-based-anal}

A quite popular analysis technique for self-suspending tasks considers the timing effects of suspensions of interfering tasks as \textit{release jitter}, while the suspension of the task under analysis is modeled as an inflation of its execution time~\cite{chen2019many}. 
Following this model, the response time $\R{i}{}$ of a self-suspending task $\SSTASK{i}$ running on core $\COREC{k}$ can be upper bounded by the least positive solution of the following recursive equation~\cite{chen2019many}:
\begin{equation}
\label{eq:analysis_self}
\R{i}{} = \WCETSHORT{i}{} + S_{i} + \sum_{\SSTASK{h} \in \HP{k}{\SSTASK{i}}} \ceilfrac{\R{i}{} + \JITs{h}{}}{\PERIOD{h}} \WCETSHORT{h}{} ,
\end{equation}
where $ \WCETSHORT{i}{}= \sum_{j=1}^{N_i^S} C_{i,j}$ and $S_{i} = \sum_{j=1}^{N_i^S-1} S_{i,j}$, while $\HP{k}{\SSTASK{i}}$ is the set of 
tasks with priority higher than $\SSTASK{i}$ allocated to $\COREC{k}$, and the value $\JITs{h}{}$ is an upper bound of the jitter induced by the overall self-suspension regions (if any) of $\SSTASK{h}$.
The value $\JITs{h}{} = \R{h}{} - \WCETSHORT{h}{}$ is a valid bound on the jitter~\cite{Nelissen2015}. If the task $\SSTASK{h}$ has no suspension region, then $\JITs{h}{} = 0$.

As previously discussed, when adopting synchronous offloading, the execution region of each accelerated segment of a task $\TASK{i}$ can be mapped 
to a suspension region. 
Thus the task model presented in this paper can be mapped to a self-suspending task, and Equation~\eqref{eq:analysis_self}
can be used to compute an upper bound of the worst-case response time.

\subsubsection{Linearizing the Jitter-based Analysis}
\label{sec:linearanalysis}

One of the advantages of the analysis technique presented above is its fitness in being linearized to be encoded in a mixed-integer linear programming (MILP) formulation.

To better understand the following steps, we first recall that a sufficient schedulability test for any task $\TASK{i}$ scheduled with a fixed-priority algorithm can be obtained by checking if there exists a value in $[0,\D{i}]$ that satisfies the following inequality~\cite{Lehoczky1989}:
\begin{equation} \label{eq:fp_suff}
\exists t \in [0,\D{i}]: W_i(t)  \le t,
\end{equation}
where $W_i(t)$ is a function bounding the overall processing time required by $\TASK{i}$ and all other tasks running in the same core that can delay $\TASK{i}$ in any time window of length $t$. Intuitively, the result follows because, if Equation~\eqref{eq:fp_suff} holds, then the processing time provided by the core (i.e., $t$) is enough to satisfy the demand of $W_i(t)$ time units.  
For the case of self-suspending tasks, considering $\SSTASK{i}$ as the task under analysis, the processor demand is expressed as (see Equation~\eqref{eq:analysis_self}): 
\begin{equation} \label{eq:wi_fp}
 W_i(t) = \WCETSHORT{i}{}  + S_{i} + \sum_{\SSTASK{h} \in \HP{k}{\SSTASK{i}}} \ceilfrac{t + \JITs{h}{}}{\PERIOD{h}} \WCETSHORT{h}{},
\end{equation}
with $\JITs{h}{} = \R{h}{} - \WCETSHORT{h}{}$ if $\SSTASK{h}$ has at least one suspension region,
and $\JITs{h}{} = 0$ otherwise. 

Pazzaglia et al.~\cite{Pazzaglia2019DATE} showed that a very accurate (less than 2\% pessimism), but sufficient schedulability test
can be obtained for a wide range of task models by just checking the inequality of Equation~\eqref{eq:fp_suff} in a limited set of points $t \in [0,\D{i}]$.
This result is particularly helpful when the test must be encoded in an MILP formulation, as it helps in drastically reducing the number of optimization variables and constraints and hence the time required to solve the optimization problem.

In the present work, the approach in~\cite{Pazzaglia2019DATE} is leveraged to encode the optimization problem aimed at finding solutions for the task-to-core and priority assignment, which optimizes the end-to-end latency of the processing chains for applications running on a heterogeneous system. 
In particular, we propose an extension of the method used in~\cite{Pazzaglia2019DATE} to handle the case of self-suspending tasks, on a per-core level. 
 
\begin{figure}[t]
\includegraphics[clip, trim=0.5cm 2cm 1.25cm 2.5cm, width = \columnwidth]{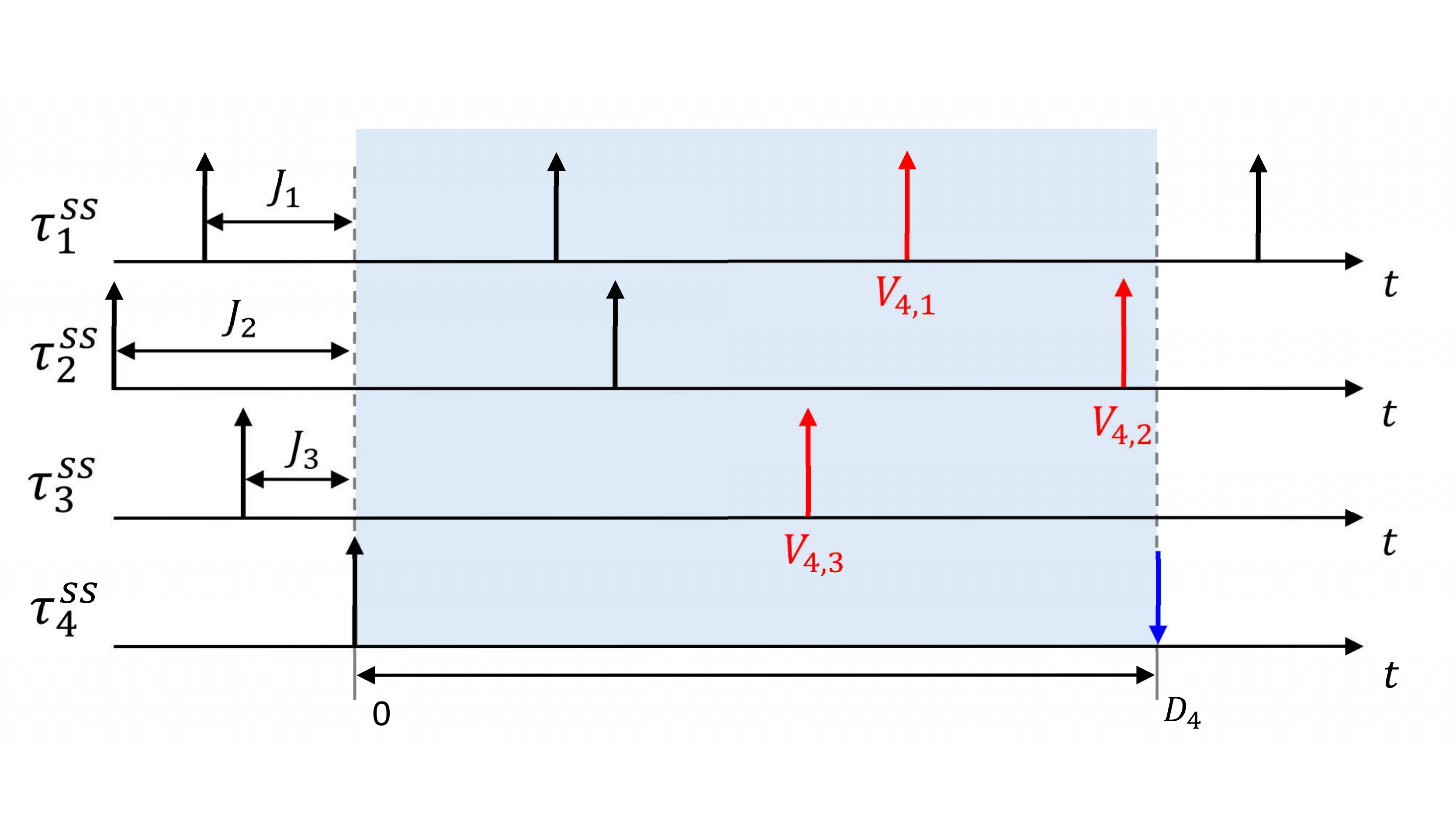}
\caption{Example of release pattern $\RELPATTERN$ with $4$ tasks. The red arrows represent the last activation of the corresponding tasks in the interval $[0,\D{4}]$ (highlighted in light blue).} 
\label{fig:pattern}
\end{figure}

The method in~\cite{Pazzaglia2019DATE} builds upon an observation first made by Park and Park~\cite{park2014efficient} according to which effective schedulability tests 
can be obtained by just checking the points in time at which the \emph{last activations} of tasks occur in the worst-case scheduling pattern of the task under analysis.
Under the jitter-based modeling of self-suspending tasks, the worst-case response time bound of Equation~\eqref{eq:analysis_self} is obtained under a release pattern defined as follows: \textbf{(i)} $\SSTASK{i}$ is released at time $t=0$ (without loss of generality) with no jitter and experiences a worst-case suspension of $S_i$ time units, \textbf{(ii)} all high-priority are ready to execute at $t=0$ after experiencing maximum jitter; and \textbf{(iii)} all successive instances are released with zero jitter~\cite{chen2019many}. We hereafter refer to this release pattern with the symbol $\RELPATTERN$, which is illustrated in Figure~\ref{fig:pattern}.
 
The following theorem provides the schedulability points of interest (following the method of~\cite{Pazzaglia2019DATE}) for a self-suspending task $\SSTASK{i}$, 
leveraging the release pattern $\RELPATTERN$.
\begin{theorem}\label{th:lastactinst}
Consider a task $\SSTASK{i} \in \SSTASKSETK{k}$ under analysis and assume that $\SSTASK{i}$ and all tasks in $\HP{k}{\SSTASK{i}}$ are released according to $\RELPATTERN$.  
A higher priority task $\SSTASK{h} \in \HP{k}{\SSTASK{i}}$ has more than one activation in $[0,\D{i}]$ if $T_h - J_h < D_i$ and its last activation in the same interval occurs at time
\begin{equation}\label{eq:Vselfsusp}
V_{i,h} = \left \lfloor \frac{D_i + J_{h}}{T_h}\right\rfloor\cdot T_h - J_{h}.
\end{equation}
\end{theorem}
\begin{proof}
By definition of $\RELPATTERN$, the first periodic instance of each task $\SSTASK{h}$ starts at time $-J_h$, as it is subject to maximum release jitter $J_h$ and it is ready to execute at time $t=0$.
Then, the second activation of each task $\SSTASK{h}$ occurs at time $T_h - J_h$. 
Hence, $\SSTASK{h}$ has more than one activation in $[0,D_i]$ if $T_h - J_h < D_i$.
If this latter condition holds, the length of the interval in which the periodic instances of $\SSTASK{h}$ overlap with the scheduling window under analysis $[0,\D{i}]$ is $D_i + J_{h}$.
Since there are $\left\lfloor (D_i + J_{h})/{T_h}\right\rfloor$ activations of $\TASK{h}$ that are fully-contained in this interval, the last one starts 
$\left\lfloor (D_i + J_{h})/{T_h}\right\rfloor T_h$ time units after the first one, which occurs at time $-J_h$.
Hence Equation~\eqref{eq:Vselfsusp} and the theorem follows.
\end{proof}

Figure~\ref{fig:pattern} shows an example where the last activation instants computed as in Theorem~\ref{th:lastactinst} are highlighted in red.

Let now $\mathcal{J}_i$ be the set of initial release jitters $J_{h}$ of each task $\SSTASK{h}\in \HP{k}{\SSTASK{i}}$  in Theorem~\ref{th:lastactinst}, i.e., 
\begin{equation}\label{eq:Jset}
\mathcal{J}_i =  \bigcup_{\SSTASK{h}\in\HP{k}{\SSTASK{i}}} J_h.
\end{equation}

The set of checkpoints $\SCHEDPOINTS{i}(\mathcal{J}_i)$ for analyzing $\SSTASK{i}$ is then obtained by the union of the points of Equation~\eqref{eq:Vselfsusp} computed for all the interfering tasks with their corresponding jitter in $\mathcal{J}_i$, plus the deadline of the task $\SSTASK{i}$, i.e.,
\begin{equation}\label{eqn:points}
\SCHEDPOINTS{i}(\mathcal{J}_i) = \bigg \{ \bigcup_{\SSTASK{h}\in\HPP{k}{\SSTASK{i}}} V_{i,h} \bigg \} \cup \left \{ D_i \right \},
\end{equation}
where $\HPP{k}{i} = \{ \SSTASK{h} \in \HP{k}{\SSTASK{i}} : T_h - J_h < D_i\}$. 
By construction, the set $\SCHEDPOINTS{i}(\mathcal{J}_i)$ contains at least one point.
Hereafter, we refer to an arbitrary check-point in $\SCHEDPOINTS{i}(\mathcal{J}_i)$ with the variable $\nu_{i,g}\in\SCHEDPOINTS{i}(\mathcal{J}_i)$.

By putting together Equation~\eqref{eq:analysis_self} and the set of check-points given by Equation~\eqref{eqn:points},
the resulting schedulability test consists in verifying the following 
condition for each task $\SSTASK{i} \in \SSTASKSETK{k}$:
\begin{equation} \label{eqn:rtpointsselfsuspensions}
\exists \nu_{i,g} \in\SCHEDPOINTS{i}(\mathcal{J}_i) ~|~  \WCETSHORT{i}{}  + S_{i} + \sum_{\SSTASK{h} \in \HP{k}{\SSTASK{i}}} \ceilfrac{\nu_{i,g} + \JITs{h}{}}{\PERIOD{h}} \WCETSHORT{h}{}  \le \nu_{i,g}.
\end{equation}

This formulation requires the knowledge of the set of jitter bounds $\mathcal{J}_i$ of each interfering task. In this work, we use  $J_{h} = \D{h} - \WCETSHORT{h}{}$ as a safe bound, which can be easily encoded in an MILP formulation.
This bound follows by noting that, since $J_{h} = \R{h}{} - \WCETSHORT{h}{}$ is a valid bound~\cite{Nelissen2015}, then under the assumption that $\D{h}{}\geq \R{h}{}$, also $\JITs{h}{} = \D{h}{} - \WCETSHORT{h}{}$ holds.
$J_{h} = \R{h}{} - \WCETSHORT{h}{}$ could provide, in principle, a less pessimistic solution. However, this choice adds a circular dependency in the response time of the target task $\SSTASK{i}$ with the response time $\R{h}{}$ of the higher priority tasks $\SSTASK{h}\in \HP{k}{\SSTASK{i}}$. 
This dependency can be broken  by introducing another set of variables (e.g., by bounding $\R{h}{}$ with a suitable checkpoint of $\SCHEDPOINTS{h}$), which has factorial complexity with respect to the number of tasks, and quickly makes the optimization impractical.
Again, note that the results presented here (which adopt the self-suspending task model) can be directly applied to the task model of this paper by leveraging the mapping of Section~\ref{sssec:mapping}.

\subsection{Response-Time Analysis for Accelerators}
\label{subsec:gpu_analysis}

As previously discussed, the length of each suspension region of a task is bounded by the maximum response time of the corresponding activity executed by the accelerator.
However, the scheduling policies used in hardware accelerators are often unknown: for example, internal details of the popular NVIDIA GPUs are not disclosed. 
Therefore, in this paper, we consider two predictable scheduling policies: \emph{round-robin} (RR) and \emph{non-preemptive fixed-priority} (NP-FP). 
Such policies can be either directly adopted by hardware accelerators, or externally enforced by application-level schedulers running on the processor cores~\cite{Bio-RTSS16, Elliot2013}  The latter can be implemented by treating the accelerator as a shared resource, thus maintaining a queue of accelerated activities (on the processing cores) and providing one activity at a time to the accelerator.

Under RR scheduling, each task executes up to one accelerated execution phase in a cyclic fashion.  
Conversely, under NP-FP, accelerated execution phases are executed to completion and scheduled according to the priorities of the corresponding tasks. 
These two policies are deemed suitable for hardware accelerators, because they allow for predictable scheduling and execution to completion (i.e., without preemptions), thus favoring cache coherence, and do not require to save and restore the context to implement preemption.

Furthermore, round-robin and NP-FP also have different interesting features. The first one ensures a fair and starvation-free access to the accelerator. In contrast, the second one guarantees  shorter delays to high-priority tasks, thus being more suitable for latency-sensitive tasks.   

By supporting multiple scheduling policies for the hardware accelerator we highlight the generality of our approach, which can also be easily extended to other scheduling policies to serve specific purposes. 

\begin{figure}[t]
	\includegraphics[width = \columnwidth]{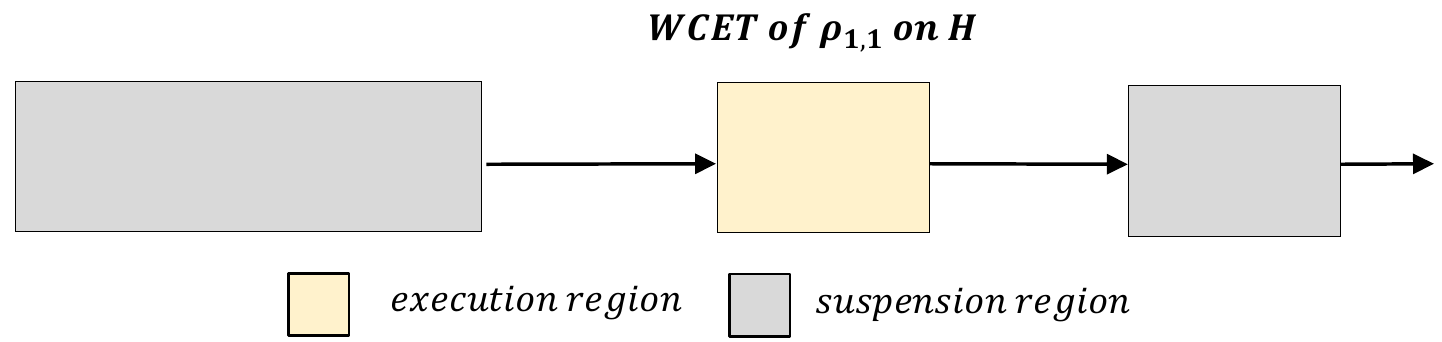}
	\caption{``Mirrored'' version of self-suspending task of Figure~\ref{fig:Model}, as it is perceived from the perspective of the hardware accelerator. }
	\label{fig:mirrored}
\end{figure} 

\smallskip
\noindent \textbf{Deriving the WCRT bounds.} 
To derive the WCRT bounds, the execution behavior of the segments running on the hardware accelerator may be modeled as an equivalent, but ``mirrored'', self-suspending task (shown in Figure~\ref{fig:mirrored}), where execution regions runs on the accelerator and suspensions regions correspond to phases running on a processor core~\cite{dong2018shared}.
However, an analysis exploiting this modeling approach would require knowing the suspension time (which in this case would occur when the task runs on the cores), as in Equation~\eqref{eq:analysis_self}, thus creating a cyclic dependency. 

To break the cycle, we compute individual upper-bounds on the suspension time of each execution phase running on the accelerator. In this way, the segment under consideration can be analyzed as a normal periodic task without self-suspension that is subject to interference due to self-suspending tasks, hence eliminating the dependency on the suspension time when deriving its WCRT bound.

\subsubsection{Round-Robin Scheduling}
\label{sec:rrbound} 

First, we consider the case in which the hardware accelerator implements a round-robin scheduling policy.

Lemma~\ref{lemma:roundrobin} proposes a bound on the suspension time of a single accelerated segment under round-robin scheduling.

\begin{lemma} \label{lemma:roundrobin}
	Consider a segment $\REGION{i}{j} \in \REGIONSET{i}^A$ of a task $\TASK{i} \in \TASKSET$, and let $l_{i,x} = \mathcal{Q}(\REGION{i}{j})$ be the corresponding suspension region of the matching task $\SSTASK{i}$.
	Then, under round-robin scheduling, the duration of $l_{i,x}$ is bounded by:
	\begin{equation}
	S_{i,x}  = \EXECR{i}{j}{\ACCELERATOR} + \sum_{\REGION{h}{u} \in \mathcal{I}(\TASK{i})} \EXECR{h}{u}{\ACCELERATOR}, 
	\end{equation}
	where $\mathcal{I}(\TASK{i})$ is the set containing, for each task $\TASK{h} \in \TASKSET \setminus \TASK{i}$, the accelerated segment $\REGION{h}{u}\in \REGIONSET{h}^A$  with longest execution phase among all segments in $\REGIONSET{h}^A$ (if $\REGIONSET{h}^A$ is not empty). 
\end{lemma}
\begin{proof}
	Due to the round-robin scheduling policy, each accelerated segment can be delayed at most once for each other segment of each other task $\TASK{h} \in \TASKSET \setminus \TASK{i}$. 
	Due to the synchronous offloading mechanism, each task $\TASK{h}$ can have at most one pending acceleration at a time. Consequently, each other task $\TASK{h} \in \TASKSET \setminus \TASK{i}$ can delay $\REGION{i}{j}$ with at most one of its accelerated segment. The lemma follows by noting that, for each $\TASK{h} \in \TASKSET \setminus \TASK{i}$, $\mathcal{I}(\TASK{i})$ contains the interfering segment with the longest execution phase. 
\end{proof}
  
Given the bound on the suspension time for an individual accelerated region, an overall bound for the suspension time of self-suspending task $\SSTASK{i}$ can be computed as $S_{i} = \sum_{\REGION{i}{j} \in \REGIONSET{i}^A}  S_{i,x}$, where $l_{i,x} = \mathcal{Q}(\REGION{i}{j})$.

This bound can be further improved by considering the ratio of periods among tasks.
The improvement is not discussed here for the sake of simplicity.

\subsubsection{Fixed-Priority Non-Preemptive Scheduling}
\label{sec:npfpanalysis}

 When the hardware accelerator provides an NP-FP scheduler, the execution phase of each accelerated segment runs on the accelerator with the same priority of the corresponding task.

Under this setting, the suspension time due to an arbitrary accelerated segment $\REGION{i}{j} \in \REGIONSET{h}^A$ is bounded by Lemma~\ref{lemma:HWA_FP}.
 
\begin{lemma} \label{lemma:HWA_FP}
		Consider a segment $\REGION{i}{j} \in \REGIONSET{i}^A$ of a task $\TASK{i} \in \TASKSET$, and let $l_{i,x} = \mathcal{Q}(\REGION{i}{j})$ be the corresponding suspension region of the matching task $\SSTASK{i}$.
		Then,  under non-preemptive fixed-priority scheduling, the duration of $l_{i,x}$ is bounded by $S_{i,x} = \Phi_{i,x} +  \EXECR{i}{j}{\ACCELERATOR}$, where $\Phi_{i,x}$ is the last positive solution of the following equation:
	\begin{equation} \label{eqn:accelnpfp}
	\Phi_{i,x} = B_{i} + \sum_{\TASK{h} \in \HPSET{\TASK{i}}{}} \ceilfrac{\Phi_{i,x} +   D_{h} - G_{h}}{\PERIOD{h}} G_h 
	\end{equation}
	where $B_{i} = \max\left\{\EXECR{l}{v}{\ACCELERATOR}\, | \, \TASK{l} \in \LPSET{\TASK{i}}{} \wedge \REGION{l}{v} \in  \REGIONSET{l}^A\right\}$, 
	$G_h = \sum_{\REGION{h}{u} \in \REGIONSET{h}^A} \EXECR{h}{u}{\ACCELERATOR}$, 
	and $\forall \TASK{h} \in \HPSET{\TASK{i}}{}$,  $R_h \leq \D{h}$ holds.
\end{lemma}
\begin{proof}
	\NEW{Since for each task $\TASK{i}$ at most one accelerated segment $\REGION{i}{j}$ can be pending at a time on $\ACCELERATOR$, as previously discussed, the duration of each suspension region $l_{i,x} = \mathcal{Q}(\REGION{i}{j})$ can be analyzed by bounding the response-time of a normal periodic task without self-suspension, with a WCET equal to $\EXECR{i}{j}{\ACCELERATOR}$, which is subject to interference due to  ``mirrored" self-suspending tasks. Such mirrored self-suspending tasks have execution regions running in the accelerator, and suspension regions that corresponds to executions on the processor cores. Interference due to high-priority tasks can be then accounted for in the analysis as for periodic tasks subject to a release jitter~\cite{chen2019many}.
	For each high-priority task, Equation~\eqref{eqn:accelnpfp} considers  the largest possible jitter (from the perspective of the execution region of $\REGION{i}{j}$ running in the accelerator), and the overall interfering WCET $G_h$ of each interfering task in the accelerator.}
	The lemma follows as a simplified instance of the jitter-based analysis for self-suspending task in~\cite{Casini2018Time} where the task under analysis has a single segment (which corresponds to the accelerated execution phase of interest), and therefore does not self-suspend.
\end{proof}

\NEW{
As for preemptive fixed-priority scheduling, the bound implied 
by Lemma~\ref{lemma:HWA_FP} can be rewritten as follows.
The duration of $l_{i,x}$ is bounded by any value $t + \EXECR{i}{j}{\ACCELERATOR}$, where $t \in [0, D_i - \EXECR{i}{j}{\ACCELERATOR}]$ (with  $l_{i,x} = \mathcal{Q}(\REGION{i}{j})$) satisfies the following inequality:
\begin{equation} \label{eq:npfp_suff}
B_{i} + \sum_{\TASK{h} \in \HPSET{\TASK{i}}{}} \ceilfrac{t + J^{A}_{h}}{\PERIOD{h}} G_h   \le t,
\end{equation}
where $J^{A}_{h} = D_{h} - G_{h}$, and $G_h$ is defined as in Lemma~\ref{lemma:HWA_FP}.

Note that Equation~\eqref{eq:npfp_suff} is similar to the ones for analyzing self-suspending provided in Section~\ref{sec:linearanalysis}.
The scheduling points derived in Theorem~\ref{th:lastactinst} and later used in Equation~\eqref{eqn:rtpointsselfsuspensions} can hence be extended to be applicable to Equation~\eqref{eq:npfp_suff}, 
provided that a suitable set $\mathcal{J}^A_{i,x} =  \bigcup_{\SSTASK{h}\in\HP{k}{\SSTASK{i}}} J^{A}_h$ of jitters for the acceleration request corresponding to each segment $l_{i,x}$ is computed. Then, the set of check-points is defined as $\SCHEDPOINTS{i}(\mathcal{J}^A_i) = \bigg \{ \bigcup_{\SSTASK{h}\in\HPP{k}{\SSTASK{i}}} V^{A}_{i,h} \bigg \} \cup \left \{ D_i \right \}$, where $V^{A}_{i,h}$ is defined as in Theorem~\ref{th:lastactinst} by replacing $J_{h}$ with $J^{A}_{h}$.

In this way, the duration of $l_{i,x}$ can be bounded by restricting the search of a value $t$ that satisfies Equation~\eqref{eq:npfp_suff} in the finite set $\SCHEDPOINTS{i}(\mathcal{J}^A_i)$.}

\section{Optimization Problem}
\label{sec:partitioning}

This section presents a mixed-integer linear programming  (MILP) formulation of the optimization problem. 
The main objectives of the proposed formulation are the following:
\begin{itemize}
	\item minimize either the end-to-end latency of the processing chains or the task WCRT bounds, according to the proposed objective function(s);
	\item select the most convenient task-to-core placement, accounting for both the change in the WCETs depending on the core type and the interference due to other tasks assigned to the same core;
	\item determine whether to accelerate tasks to find the most convenient trade-off between shorter WCETs occurring when a task is accelerated and longer delays in the accelerator when many activities are offloaded;
	\item optimize the priority assignment; and
	\item ensure schedulability, i.e., guarantee that each task always completes within its deadline.
\end{itemize}

\NEW{
Section~\ref{sec:var} introduces the main variables of the problem, which do not depend on the scheduling policy adopted on the accelerator.
Section~\ref{sec:mainconstraints} presents the main constraints regarding task-to-core mapping, priority and accelerated segments. Next, Section~\ref{ssec:milpwcetbounds} lists the constraints to bound the WCETs of tasks in the processors, while Section~\ref{ssec:milpwcrtbounds} presents the constraints to bound the WCRTs of the tasks. Sections~\ref{sec:rrconstraints} and~\ref{sec:npfpconstraints} present the set of variables and the constraints at the accelerator level for RR and NP-FP scheduling policies, respectively. Only one of them needs to be used in an optimization problem instance, depending on the scheduling policy adopted for the accelerator.}
This gives rise to a modular approach that may be extended to other scheduling policies by just introducing a new set of variables and constraints, while leaving most of the optimization problem unaltered.
Finally, Section~\ref{sec:objectivefunction} presents different objective functions that can be used in the optimization problem. 

In the MILP constraints we often leverage the so-called \emph{big-M} formulation: to this end, we define the symbol $\BIGM$, a very large positive constant (representing infinity).

\subsection{Main MILP Variables} 
\label{sec:var}

We start presenting the main variables needed to describe the problem.
Other auxiliary and additional variables are introduced when required.

\subsubsection{Boolean Variables}
\label{sec:booleanvar}
\begin{itemize}
	\item \emph{Task assignment in core:} For each task $\TASK{i} \in \TASKSET$, and for each core $\COREC{k} \in \PROCESSORSET$, $\PROCVAR{i}{k} \in \{0,1\}$ is a binary variable set to $1$ if $\TASK{i}$ is allocated to $\COREC{k}$; $0$ otherwise.

	\item \emph{Tasks in same core:} For each task pair $\TASK{i}, \TASK{s}  \in \TASKSET$, with $i \neq s$, $\SAMEPROC{i}{s} \in \{0,1\}$ is set to $1$ if  $\TASK{i}$ is allocated on the same core as $\TASK{s}$; $0$ otherwise.
	\item \emph{Task priority assignment:}  For each task $\TASK{i} \in \TASKSET$, for each $q \in  \mathbb{N}$, $1 \leq q \leq |\TASKSET|$, $\PRIO{i}{q}\in \{0,1\}$ is equal to $1$ if $\TASK{i}$ is assigned priority $q$; $0$ otherwise.
	\item \emph{Priority relationship between tasks:} For each task pair $\TASK{i}, \TASK{s}  \in \TASKSET$, with $i \neq s$, $\HIGHERPRIO{i}{s}\in \{0,1\}$ is equal to $1$ if task $\TASK{i}$ is assigned a higher priority than $\TASK{s}$; $0$ otherwise.

	\item \emph{Accelerated segment:} For each task $\TASK{i} \in \TASKSET$, and for each segment $\REGION{i}{j}\in\REGIONSET{i}$, $\ACV{i,j}  \in \{0,1\}$ is set to $1$ if and only if $\REGION{i}{j}$ is offloaded to the accelerator; $0$ otherwise.
	\item \emph{Selector variable for candidate WCRT of a task:} For each $\TASK{i} \in \TASKSET$, for each $\nu_{i,g} \in \SCHEDPOINTS{i}$, $\ENVAR{i,g}  \in \{0,1\}$ is a binary variable set to 1 if $\nu_{i,g}$ is the candidate WCRT bound. 
\end{itemize}	
	
\subsubsection{Real and Integer Variables} 
\label{sec:realintvar}
\begin{itemize}

	\item  \emph{Priority index of a task:} For each task $\TASK{i} \in \TASKSET$, $\PRIOABS{i}\in\mathbb{N}^{> 0}$ encodes the absolute priority of $\TASK{i}$.
  	\item \emph{WCETs of tasks and segments:} For each task $\TASK{i} \in \TASKSET$, and for each segment $\REGION{i}{j} \in \REGIONSET{i}$, $ \ETASK{i}\in\mathbb{R}^{\geq 0}$ and $\ESEGM{i}{j}\in\mathbb{R}^{\geq 0}$ are the WCET of $\TASK{i}$ and $\REGION{i}{j}$, respectively, depending on the core where it is allocated and on whether it is accelerated or not.  

	\item  \emph{WCET of an interfering task:} For each task pair $\TASK{i}, \TASK{s}  \in \TASKSET$, with $i \neq s$, $\INTERFWCET{i}{s}\in\mathbb{R}^{\geq 0}$ is equal to the WCET of $\TASK{i}$ (on the core where it is allocated) if it can interfere with $\TASK{s}$; $0$ otherwise.
	\item  \emph{Response time candidate of a task:} For each $\TASK{i} \in \TASKSET$, for each $\nu_{i,g} \in \SCHEDPOINTS{i}$ (see Equation~\eqref{eqn:points}), $\RTCVAR{i,g} \in \mathbb{R}^{\geq 0}$ is a candidate WCRT bound for $\TASK{i}$. 
	\item \emph{Response time of a task:} For each $\TASK{i} \in \TASKSET$,  $\RTVAR{i} \in \mathbb{R}^{\geq 0}$ is the WCRT bound of $\TASK{i}$.
\end{itemize}	
	
\subsubsection{Common Variables for Hardware Acceleration} 
  \label{sec:hwaccelerationvar}
  
\begin{itemize}
\item \emph{Suspension time of a segment:} For each task $\TASK{i} \in \TASKSET$, for each segment $\REGION{i}{j} \in \REGIONSET{i}$, $\SR{i,j}\in\mathbb{R}^{\geq 0}$ bounds the time spent by $\REGION{i}{j}$ on the hardware accelerator.  
\item \emph{Suspension time of a task:} For each task $\TASK{i} \in \TASKSET$, $\ST{i}\in\mathbb{R}^{\geq 0}$ bounds the overall time spent by $\TASK{i}$ on the hardware accelerator.  
\end{itemize}

\subsection{Basic Mapping Constraints} 
\label{sec:mainconstraints}

 First, we enforce each task to be assigned to only one processor, through the variable $ \PROCVAR{i}{k}$.
 \begin{constraint} [Task-to-core mapping] \label{constraint:oneprocessor}
 	For each $\TASK{i} \in \TASKSET$, 
	$$ 	\sum_{\COREC{k} \in \PROCESSORSET} \PROCVAR{i}{k} = 1.	$$
 \end{constraint}
 \begin{proof}
 	By definition, $\PROCVAR{i}{k}$ is set to 1 if and only if $\TASK{i}$ is assigned to core $\COREC{k}$. The constraint follows noting that $\TASK{i}$ is allocated to only one processor only if this constraint is imposed.
 \end{proof}

It is convenient to introduce some auxiliary variables to cope with the task-to-processor assignment:
\begin{itemize}
\item \emph{Tasks in same core $\COREC{k}$:} For each task pair $\TASK{i}, \TASK{s}  \in \TASKSET$, with $i \neq s$, for each core $\COREC{k} \in \PROCESSORSET$, $\SAMEPROCK{i}{s}{k} \in \{0,1\}$ is set to $1$ if $\TASK{i}$ and $\TASK{s}$ are both allocated onto $\COREC{k}$; $0$ otherwise.
\end{itemize}

Constraint~\ref{constraint:sameproc} enforces the definition of variables $\SAMEPROCK{i}{s}{k}$ and $\SAMEPROC{i}{s}$, which denote if two tasks are in the same processor.
 
 \begin{constraint} [Tasks in the same core] \label{constraint:sameproc}
For each task pair $\TASK{i}, \TASK{s}  \in \TASKSET$, with $i \neq s$, and for each core $\COREC{k} \in \PROCESSORSET$,
$$
\SAMEPROCK{i}{s}{k} \ge 1 - (2 - \PROCVAR{i}{k} - \PROCVAR{s}{k}), $$
$$ \SAMEPROCK{i}{s}{k} \le \PROCVAR{i}{k}, \quad \SAMEPROCK{i}{s}{k} \le \PROCVAR{s}{k}.$$
Then, for each pair of tasks $\TASK{i} \in \TASKSET$, $\TASK{s} \in \TASKSET \setminus \TASK{i}$:
$$
\SAMEPROC{i}{s} = 	\sum_{\COREC{k} \in \PROCESSORSET} \SAMEPROCK{i}{s}{k}.
$$
\end{constraint}
\begin{proof}
	By definition,  $\SAMEPROCK{i}{s}{k} \in \{0,1\}$ is set to $1$ if and only if task $\TASK{i}$ is allocated on the same core $\COREC{k} \in \PROCESSORSET$ of task $\TASK{s}$.
	If $\PROCVAR{i}{k} = \PROCVAR{s}{k} = 1$, by substituting in the constraint $\SAMEPROCK{i}{s}{k} \ge 1 \wedge \SAMEPROCK{i}{s}{k} \le 1 \Rightarrow \SAMEPROCK{i}{s}{k} = 1$ is enforced. 
	If $\PROCVAR{i}{k} = \PROCVAR{s}{k} = 0$ or $\PROCVAR{i}{k} \neq \PROCVAR{s}{k}$, by substituting in the constraint we get $\SAMEPROCK{i}{s}{k} \le 0$ for at least one of the last two inequalities, 
while the first inequality enforces either $\SAMEPROCK{i}{s}{k} \ge -1$ or $\SAMEPROCK{i}{s}{k} \ge 0$. This implies $\SAMEPROCK{i}{s}{k} = 0$, proving the first set of constraints.
Finally, since due to Constraint~\ref{constraint:oneprocessor} each task is assigned to a core, $\sum_{\COREC{k} \in \PROCESSORSET} \SAMEPROCK{i}{s}{k}$ can be either zero or one, and the last equality enforces the definition of variable $\SAMEPROC{i}{s}$.
\end{proof}
 
 Constraint~\ref{constraint:uniquepriority_one} enforces the uniqueness of the priority assignment through the boolean variable $\PRIO{i}{p}$.
 \begin{constraint} [Uniqueness of the priority] \label{constraint:uniquepriority_one} 
 	
 	For each task $\TASK{i} \in \TASKSET$, 
 	$$\sum_{p \in \{1, \ldots, |\TASKSET|\}} \PRIO{i}{p} = 1,$$
	and for each priority $p \in \{1, \ldots, |\TASKSET|\}$,
	$$\sum_{\TASK{i} \in \TASKSET} \PRIO{i}{p} = 1$$
\end{constraint}
 \begin{proof}
	By definition, $\PRIO{i}{p}$ is set to $1$ if and only if $\TASK{i}$ is assigned to priority $p$. The constraint follows noting that the uniqueness of the priority holds only if \textbf{(i)} each task $\TASK{i} \in \TASKSET$ is assigned to exactly one priority $p$ ($p \in \{1, \ldots, |\TASKSET|$), and \textbf{(ii)}  each priority $p$ ($p \in \{1, \ldots, |\TASKSET|$) is assigned to exactly one task $\TASK{i} \in \TASKSET$. 
\end{proof}

Constraint~\ref{constraint:priorityvalue} specifies the value of the integer variable $\PRIOABS{i}$, encoding the absolute priority of $\TASK{i}$.
\begin{constraint} [Priority index of a task] \label{constraint:priorityvalue} 
	For each task $\TASK{i} \in \TASKSET$,   $$\PRIOABS{i} = \sum_{1 \leq p \leq|\TASKSET|} p \cdot \PRIO{i}{p}.$$
\end{constraint} 
 \begin{proof}
	By definition, $\PRIOABS{i}$ is an integer value encoding the absolute priority of a corresponding task  $\TASK{i} \in \TASKSET$. By Constraint~\ref{constraint:uniquepriority_one}, $\PRIO{i}{p}$ is set to $1$ if $\TASK{i}$ is assigned to priority $p$, and there is only one task with such a priority. The constraint follows by summing up the priority index multiplied by the boolean variable $\PRIO{i}{p}$, with $p \in \{1, \ldots, |\TASKSET| \}$. 
\end{proof}

Next, we limit the possible values of $\HIGHERPRIO{i}{s}$. This is done with the combination of two constraints. With Constraint~\ref{constraint:higherpriorityonlyoneway} we enforce that, for each pair of tasks $\TASK{i}, \TASK{s} \in \TASKSET$, either $\TASK{i}$ has higher priority than $\TASK{s}$ or vice-versa.

\begin{constraint} [Priority relationship between tasks] \label{constraint:higherpriorityonlyoneway}
	For \\ each pair of tasks $\TASK{i} \in \TASKSET$, $\TASK{s} \in \TASKSET \setminus \TASK{i}$,
	$$ \HIGHERPRIO{i}{s} + \HIGHERPRIO{s}{i}= 1.$$
\end{constraint}
\begin{proof}
	By definition, $\HIGHERPRIO{i}{s}\in \{0,1\}$ is equal to $1$ if task $\TASK{i}$ is assigned to a higher priority than $\TASK{s} \in \TASKSET \setminus \TASK{i}$. The constraint enforces that either $\TASK{i} \in \TASKSET$ has higher priority than $\TASK{s}$, or vice versa, by imposing that only one between $\HIGHERPRIO{i}{s}$ and $\HIGHERPRIO{s}{i}$ can be set to one. 
\end{proof}

Secondly, Constraint~\ref{constraint:relativepriorities} enforces the relationship between absolute priorities, encoded by variables $\PRIOABS{i}$, and relative priorities, encoded by variables $\HIGHERPRIO{i}{s}$. 
\begin{constraint} [Relative and absolute priorities] 
	
	\label{constraint:relativepriorities} 
	
	For each pair of tasks $\TASK{i} \in \TASKSET$, $\TASK{s} \in \TASKSET \setminus \TASK{i}$,
	$$-\HIGHERPRIO{i}{s}\cdot \BIGM \le \PRIOABS{s} - \PRIOABS{i} \le (1 - \HIGHERPRIO{i}{s})\cdot \BIGM. $$

\end{constraint}
 \begin{proof} 

 	 If $\HIGHERPRIO{i}{s} = 0$, then we have $0 \le \PRIOABS{s} - \PRIOABS{i} \le \BIGM $. Hence $0 \le \PRIOABS{s} - \PRIOABS{i}$, and  $\PRIOABS{i} < \PRIOABS{s}$ because Constraint~\ref{constraint:uniquepriority_one} enforces the uniqueness of the priority assignment. 

 	 If $\HIGHERPRIO{i}{s} = 1$, then we have $-M \le \PRIOABS{s} - \PRIOABS{i} \le 0 $. Hence $\PRIOABS{s} - \PRIOABS{i} \le 0$, and  $\PRIOABS{s} < \PRIOABS{i}$. 
 	 The constraint follows.

 \end{proof}

Finally, Constraint~\ref{constraint:accelerationpossible} specifies that, for each task $\TASK{i} \in \TASKSET$ and for each segment $\REGION{i}{j}$ the processing phase can be executed on a CPU only if $\CANACCEL{i,j} \neq \mbox{\texttt{HWA}}$, and a segment is executed on the accelerator if $\CANACCEL{i,j} = \mbox{\texttt{HWA}}$, constraining the possible values of $\ACV{i,j}$.  

 \begin{constraint} [HW Acceleration] \label{constraint:accelerationpossible}
 	
  	For each task $\TASK{i} \in \TASKSET$, for each segment $\REGION{i}{j}\in\REGIONSET{i}$, if $\CANACCEL{i,j} = \mbox{\texttt{CPU}}$, then 
 	$$ \ACV{i,j} = 0.$$
 	If $\CANACCEL{i,j} = \mbox{\texttt{HWA}}$ then 
 	$$ \ACV{i,j} = 1.$$ 
 \end{constraint}
\begin{proof}
	By definition, $\ACV{i,j}  \in \{0,1\}$ is set to $1$ if and only if $\REGION{j}{i}$ is offloaded to the accelerator, and it is set to $0$ otherwise.
	If $\CANACCEL{i,j} = \mbox{\texttt{CPU}}$, then no HW-accelerated implementation of $\REGION{i}{j}$ is available: hence, $\ACV{i,j} = 0$ is enforced. 
	Conversely, if $\CANACCEL{i,j} = \mbox{\texttt{HWA}}$, only an HW-accelerated implementation of $\REGION{j}{i}$ is available, and hence $\ACV{i,j} = 1$ is imposed. The constraint follows.  
\end{proof}

\subsection{Bounding WCETs and CPU Interference}
\label{ssec:milpwcetbounds}

 Next, we present a set of constraints to characterize the WCET of the segments and tasks.
The following constraints will act only as safe lower bounds for the variables related to WCETs. This approach limits the complexity of the MILP formulation, and is justified by the fact that, whenever the objective function is a minimization involving the response time, the solver chooses the smallest possible value allowed by the constraints, for all the variables that contribute to the response time. 
Additionally, the WCET values chosen by the solver will also be limited by the necessity of guaranteeing the schedulability of the system (Constraint~\ref{constraint:schedulability}).
 
 Constraint~\ref{constraint:executiontimeregion} 
imposes the value of $\ESEGM{i}{j}$ to be no smaller than the WCET of the corresponding segment $\REGION{i}{j}$ on the core $\COREC{k}$ where the MILP solver allocated $\TASK{i}$, 
which depends on whether $\REGION{i}{j}$ is accelerated or not.

\begin{constraint} [WCET of a segment]
	 
	 \label{constraint:executiontimeregion} 
	
	For each task $\TASK{i} \in \TASKSET$, for each segment $\REGION{i}{j} \in \REGIONSET{i}$, 
	\begin{align*}
	&\ESEGM{i}{j} \ge \sum_{\COREC{k} \in \PROCESSORSET}  \EXECR{i}{j}{\COREC{k}}\cdot \PROCVAR{i}{k}  - \ACV{i,j} \cdot \BIGM \\
	&\ESEGM{i}{j} \ge \sum_{\COREC{k} \in \PROCESSORSET}  w_{i,j}(\COREC{k})  \cdot \PROCVAR{i}{k} - (1 - \ACV{i,j}) \cdot \BIGM,
	\end{align*}
	where $w_{i,j}(\COREC{k})  = \OFFR{i}{j}{\COREC{k}} + \FINR{i}{j}{\COREC{k}}$. 
\end{constraint}
\begin{proof} 
	Recall that the variable $\ESEGM{i}{j}$ denotes the WCET of $\REGION{i}{j}$, depending on the core where it is allocated and on whether it is accelerated or not. 	
    If $\REGION{i}{j}$ is not accelerated, then $\ACV{i,j} = 0$, hence   
    $\ESEGM{i}{j} \ge \sum_{\COREC{k} \in \PROCESSORSET}  \EXECR{i}{j}{\COREC{k}}\cdot \PROCVAR{i}{k} $,   
    while the second inequality does not have effect (i.e., since $\BIGM$ is a large constant, it is reduced to $\ESEGM{i}{j} \ge -\infty$). By Constraint~\ref{constraint:oneprocessor}, $\PROCVAR{i}{k}$ is equal to $1$ only for the core $\COREC{k}$ where $\TASK{i}$ is allocated, hence the sum 
     $\sum_{\COREC{k} \in \PROCESSORSET}  \EXECR{i}{j}{\COREC{k}} \cdot \PROCVAR{i}{k}$ 
      actually consists in only one term, while the others are set to $0$. Such a term is the WCET of $\TASK{i}$ in the core where it is allocated, which consists of only the processing phase, with WCET $\EXECR{i}{j}{\COREC{k}}$. 
    On the other hand, if $\REGION{i}{j}$ is accelerated, $\ACV{i,j} = 1$, hence the first inequality has no effect, while the second becomes
    $\ESEGM{i}{j} \ge \sum_{\COREC{k} \in \PROCESSORSET} w_{i,j}(\COREC{k}) \cdot \PROCVAR{i}{k}$. Since each term in the sum is multiplied by $\PROCVAR{i}{k}$ as in the previous case, $\ESEGM{i}{j}$ is constrained to be greater than or equal to the WCET of the offloading and finalization phases of $\REGION{i}{j}$ (i.e., $w_{i,j}(\COREC{k})$), referred to the specific core $\COREC{k}$ where the task $\TASK{i}$ is mapped. The constraint follows.  
\end{proof}

Constraint~\ref{constraint:executiontimetasks} enforces the definition of the variables $\ETASK{i}$.  
\begin{constraint} [WCET of a task]
	
	\label{constraint:executiontimetasks} 
	For each task $\TASK{i} \in \TASKSET$, 
	$$ \ETASK{i} \ge \sum_{\REGION{i}{j} \in \REGIONSET{i}} \ESEGM{i}{j}$$	
\end{constraint}
\begin{proof} 
 	The constraint follows by noting that the overall WCET of $\TASK{i} \in \TASKSET$ is the sum of the individual WCETs of each segment $\REGION{i}{j} \in \REGIONSET{i}$. 
\end{proof}

Constraint~\ref{constraint:interferenceWCETtotask} copes with variable $\INTERFWCET{i}{s}$, which bounds the interference of one job of task $\TASK{s}$ on a job of $\TASK{i}$. The value of $\INTERFWCET{i}{s}$ is constrained to be greater than or equal to $\ETASK{s}$, if $\TASK{s} \in \TASKSET \setminus \TASK{i}$ can interfere with $\TASK{i} \in \TASKSET$, zero otherwise.

\begin{constraint} [CPU Interference]
	
	 \label{constraint:interferenceWCETtotask} 
	
	For each pair of tasks $\TASK{i} \in \TASKSET$, $\TASK{s} \in \TASKSET \setminus \TASK{i}$,
	$$\INTERFWCET{i}{s} \ge \ETASK{i} - \BIGM \cdot (2 - \HIGHERPRIO{i}{s} - \SAMEPROC{i}{s})$$
\end{constraint}  
 
 \begin{proof} 
 	Under a partitioned fixed-priority scheme, a task  $\TASK{i}$ may interfere with another task $\TASK{s}$ if and only if: \textbf{(i)} it is allocated in the same core, and \textbf{(ii)} it has higher priority than $\TASK{s}$.
 	Condition (i) is verified when $\SAMEPROC{i}{s} = 1$ (Constraint~\ref{constraint:sameproc}), whereas condition (ii) holds when $\HIGHERPRIO{i}{s} = 1$  (Constraint~\ref{constraint:relativepriorities}).  
	The constraint follows by noting that $\INTERFWCET{i}{s} \ge \ETASK{i}$ is enforced if and only if $\SAMEPROC{i}{s} = 1$ and $\HIGHERPRIO{i}{s} = 1$. Otherwise, the constraint has no effect ($\INTERFWCET{i}{s} \ge -\infty$). 
\end{proof}  

\subsection{Bounding WCRTs}
\label{ssec:milpwcrtbounds}

As discussed in Section~\ref{sec:linearanalysis}, the processor demand constraint that uses the set of checkpoints in Theorem~\ref{th:lastactinst} is a suitable way to bound the WCRTs in a linear optimization problem. This however requires the knowledge of the release jitter for each interfering self-suspending task. The choice of a jitter $J_{h} = \D{h} - \WCETSHORT{h}{}$ for $\TASK{h}$ is a safe bound for a self-suspending task. 
However, such jitter still depends on the WCET of the task. In particular, due to the heterogeneity of the platform, the WCET depends on \textbf{(i)} the type of the processor where it is allocated, and \textbf{(ii)} whether it is accelerated. This way, it would be an explicit function of $\PROCVAR{i}{k}$ and $\ACV{i,j}$. 
As a consequence, we would be required to introduce additional variables to compute the floor term of Equation~\eqref{eqn:points} and the ceiling term of Equation~\eqref{eqn:rtpointsselfsuspensions}. Additionally, the ceiling term of Equation~\eqref{eqn:rtpointsselfsuspensions} is multiplied by the WCET of each interfering task, which is itself a variable (i.e., $\INTERFWCET{j}{i}$), thus making the problem not linear. 

To address this problem, we consider a more conservative (but linear) approach. 
Since the response-time bound is monotonic non-decreasing with respect to the jitter bound, which in turn is monotonic non-increasing with the WCET of the interfering task, by increasing the jitter of the interfering tasks we obtain a more conservative estimate of the WCRT of the task under analysis. Hence, taking the minimum WCET over all possible configurations of processor type and acceleration state of each segment yields a safe bound on the jitter. 
To this end, we introduce the \emph{constant} term $\WCETMIN{i}$ for a task $\TASK{i}$ to denote the \emph{minimum} WCET with which a task can be characterized, among all possible configurations of cores $\COREC{k} \in \PROCESSORSET$ and all the possible combinations of accelerated (or not) segments $\REGION{i}{j}\in\REGIONSET{i}$. 
Additionally, for each task $\TASK{i}\in \TASKSET$, since the priority level is a variable of the optimization problem, we do not know in advance which tasks will have higher priority than $\TASK{i}$, thus we consider the (eventual) checkpoint associated to each task in the set. 
Note that extending the approach of Section~\ref{sec:linearanalysis} to all tasks in $\TASKSET$ (compared with only the ones that have higher priority than $\TASK{i}$) has the only effect of possibly introducing additional checkpoints to the schedulability test of $\TASK{i}$, which the solver is required to check.

By considering all tasks $\TASK{s}\in\TASKSET\setminus\TASK{i}$ as potentially having higher priority than $\TASK{i}$, 
the jitters used in Section~\ref{s:jitter-based-anal} can be refined as follows:
\begin{equation}\label{eq:jitterMILP}
\overline{\mathcal{J}}_i =  \bigcup_{\TASK{s}\in\TASKSET\setminus\TASK{i}} \left\{  J_s\right\},
\end{equation} 
with 
\begin{equation}\label{eq:jitterMILP1}
J_s = \left\{
\begin{array}{lc}
 \DEADLINE{s} - \WCETMIN{s} & \mbox{if } \REGIONSET{s}^H \neq \emptyset \\
0 & \mbox{otherwise,}
\end{array}\right.
\end{equation} 
which is used to build the set of checkpoints $\SCHEDPOINTS{i}(\overline{\mathcal{J}}_i)$ as:
\begin{equation}\label{eq:checkpointsMILP}
\SCHEDPOINTS{i}(\overline{\mathcal{J}}_i) = \left\{\bigcup_{\TASK{s}\in\TASKSET\setminus\TASK{i}} V_{i,s}\right\} \cup \{ D_i \},
\end{equation}
where $V_{i,s}$ is defined as in Theorem~\ref{th:lastactinst}.

   Constraint~\ref{constraint:WCRTbound} establishes a response-time bound for each task $\TASK{i} \in \TASKSET$ with the method presented in Section~\ref{sec:linearanalysis}, and considering the jitter set computed as in Equation~\eqref{eq:jitterMILP}. 

\begin{constraint} [WCRT bound candidate] \label{constraint:WCRTbound}
	For each task  \\ $\TASK{i} \in \TASKSET$, and for each $\nu_{i,g} \in \SCHEDPOINTS{i}(\overline{\mathcal{J}}_i)$ obtained with Equation~\eqref{eq:checkpointsMILP} using the jitter set of Equation~\eqref{eq:jitterMILP},
	\begin{align*}
	&\RTCVAR{i,g} \ge \ETASK{i} + \ST{i} + \sum_{\TASK{s} \in \TASKSET \setminus \TASK{i} } \ceilfrac{\nu_{i,g} + J_s}{\PERIOD{s}} \INTERFWCET{s}{i},\\
	&\RTCVAR{i,g} \le \nu_{i,g} + (1 - \ENVAR{i,g}) \cdot \BIGM, \\
	&\RTVAR{i} \ge \RTCVAR{i,g} - (1 - \ENVAR{i,g}) \cdot \BIGM.
	\end{align*}
		Additionally, for each task $\TASK{i} \in \TASKSET$, 	
	$$\sum_{\nu_{i,g} \in \SCHEDPOINTS{i}(\overline{\mathcal{J}}_i)} \ENVAR{i,g} = 1.$$  
\end{constraint}
\begin{proof} 
	For each $\TASK{i} \in \TASKSET$, there exists up to  $|\SCHEDPOINTS{i}|$ candidate response-time bounds, represented with variables $\RTCVAR{i,g}$, where the index $g$ corresponds to the index of $\nu_{i,g} \in \SCHEDPOINTS{i}(\overline{\mathcal{J}}_i)$.

	For each $\nu_{i,g} \in \SCHEDPOINTS{i}(\overline{\mathcal{J}}_i)$, the first inequality implements the response-time bound of Equation~\eqref{eq:analysis_self} 
	bounding the jitter as $\JITs{s}{} = \DEADLINE{s} - \WCETMIN{s}$.
	The second inequality enforces that the selected response-time bound must be valid according to Equation~\eqref{eqn:rtpointsselfsuspensions}. In all other cases, the inequality is disabled (i.e., $\RTCVAR{i,g} \le \infty$). 
	The third inequality enforces that the response-time bound is greater than or equal to the selected response-time candidate: otherwise, it is disabled (i.e., $\RTCVAR{i,g} \ge -\infty$). 
		Finally, the fourth inequality enforces only one of the checkpoints to be selected as the actual response-time bound.
\end{proof}

Finally, Constraint~\ref{constraint:schedulability} enforces $\RTVAR{i}$ to be a valid response-time bound.
\begin{constraint}[Schedulability] \label{constraint:schedulability}
	For each task $\TASK{i} \in \TASKSET$, 
	
	$$ \RTVAR{i} \le \DEADLINE{i}.$$ 
\end{constraint}

Next, we show how to bound the suspension time spent in the hardware accelerator. Clearly, this depends on the specific scheduling policy implemented by the accelerator. We present the cases in which such policies are round-robin and non-preemptive fixed priority, as a set of constraints that need to be added to the optimization problem \emph{only when the scheduling policy is used}. In this way, we highlight the modular nature of our approach, which can easily be extended to other scheduling policies by just implementing some new constraints, while most of the optimization problem can be left unaltered. 

\subsection{Worst-Case Suspension Time with RR}
\label{sec:rrconstraints}
 
This section presents the constraints to bound the duration of accelerated segment (and hence of the suspension time of the task running on the core) under round-robin scheduling.

Before proceeding, it is necessary to introduce additional variables to handle the RR scheduling.

\subsubsection{Additional Variable for RR Scheduling} 
  \label{sec:rrvar}
\begin{itemize}
\item \emph{Longest accelerated segment:} For each task $\TASK{i} \in \TASKSET$, $\LA{i}\in\mathbb{R}^{\geq 0}$ bounds the longest WCET of an accelerated segment $\REGION{i}{j} \in \ACCREGIONSET{i}$  of task $\TASK{i}$.  
\end{itemize}

\subsubsection{Constraints}

First, Constraint~\ref{constraint:acceleratedregionlongest} bounds the length of the longest accelerated segment of each task $\TASK{i} \in \TASKSET$, enforcing the definition of $\LA{i}$.

\begin{constraint}[Longest accelerated segment] \label{constraint:acceleratedregionlongest} 
	For each task $\TASK{i} \in \TASKSET$, and for each segment $\REGION{i}{j} \in \REGIONSET{i}^H$, 
	$$ \LA{i} \ge \EXECR{i}{j}{\ACCELERATOR} - (1 - \ACV{i,j}) \cdot \BIGM $$
\end{constraint}
\begin{proof} 
	When a segment $\REGION{i}{j} \in \REGIONSET{i}^H$ is accelerated, $\EXECR{i}{j}{\ACCELERATOR}$ bounds its WCET. 
	The constraint follows by noting that if $\REGION{i}{j}$ is accelerated, $\ACV{i,j} = 1$ and $\LA{i} \ge \EXECR{i}{j}{\ACCELERATOR}$ is imposed; otherwise, $\ACV{i,j} = 0$ and the constraint does not take effect (i.e. $\LA{i} \ge -\infty$).
\end{proof}

Constraint~\ref{constraint:accelerationtimeregion} bounds the time spent on the hardware accelerator by each segment $\REGION{i}{j} \in \REGIONSET{i}^H$.
\begin{constraint}[Suspension time of a segment] \label{constraint:accelerationtimeregion} 
	For each \\ task $\TASK{i} \in \TASKSET$, and for each segment $\REGION{i}{j} \in \REGIONSET{i}^H$, 
	$$ \SR{i,j} \ge \EXECR{i}{j}{\ACCELERATOR} + \bigg(\sum_{\TASK{j} \in \TASKSET \setminus \TASK{i}} \LA{j} \bigg)  - (1 - \ACV{i,j}) \cdot \BIGM $$
\end{constraint}
\begin{proof} 
   The constraint follows from Lemma~\ref{lemma:roundrobin}.
\end{proof}

Constraint~\ref{constraint:accelerationtimetask} bounds the time spent on the hardware accelerator by each task $\TASK{i} \in \TASKSET$.
\begin{constraint}[Suspension time of a task]
	\label{constraint:accelerationtimetask} 
	For each task $\TASK{i} \in \TASKSET$ such that $\REGIONSET{i}^H \neq \emptyset$,  
	$$ \ST{i} \ge \sum_{\REGION{i}{j} \in \REGIONSET{i}} \SR{i,j} $$
\end{constraint}
\begin{proof} 
	The constraint follows by noting that the overall time spent on the hardware accelerator by each task $\TASK{i}$ is bounded by the time spent on the hardware accelerator by each segment $\REGION{i}{j} \in \REGIONSET{i}$, encoded by variables $\SR{i,j}$.
\end{proof}

\subsection{Worst-Case Suspension Time with NP-FP}
\label{sec:npfpconstraints}

This section presents the constraints to bound the time globally spent on the hardware accelerator, and the overall suspension-time experienced by the tasks running on the cores under non-preemptive fixed priority scheduling.  

Before proceeding, it is necessary to introduce additional variables to handle the
NP-FP scheduling.

\subsubsection{Additional Variables for NP-FP Scheduling} 
\label{sec:npfpvar}

\begin{itemize}
	\item  \emph{WCET of an interfering segment on $\ACCELERATOR$:} 
	For each task $\TASK{i} \in \TASKSET$, for each segment $\REGION{i}{j} \in \ACCREGIONSET{i}$, for each task $\TASK{s}  \in \TASKSET \setminus \TASK{i}$, $\EACCRVAR{i}{j}{s}\in\mathbb{R}^{\geq 0}$  is equal to the WCET of $\REGION{i}{j}$ (on $\ACCELERATOR$) if $\TASK{i}$ can interfere with $\TASK{s}$; it is $0$ otherwise.  
	\item  \emph{WCET of an interfering task on $\ACCELERATOR$:} 
	For each task pair $\TASK{i}, \TASK{s} \in \TASKSET$, with $i \neq s$, $\EACCVAR{i}{s}\in\mathbb{R}^{\geq 0}$  is equal to the WCET of $\TASK{i}$ (on $\ACCELERATOR$) if it can interfere with $\TASK{s}$; $0$ otherwise.  
	\item  \emph{NP Blocking time of a task on $\ACCELERATOR$:} 
	For each task $\TASK{i} \in \TASKSET$, $\BLOCKVAR{i} \in\mathbb{R}^{\geq 0}$ bounds the blocking due to lower-priority tasks that any segment of $\TASK{i}$ can experience when accelerated on $\ACCELERATOR$ due to non-preemptive scheduling. 
	\item  \emph{Suspension time candidate of a segment:} For each $\TASK{i} \in \TASKSET$, for each segment $\REGION{i}{j} \in \ACCREGIONSET{i}$, for each $\nu_{i,g} \in \SCHEDPOINTS{i}$, $\STCVAR{i,j,g} \in \mathbb{R}^{\geq 0}$ is a candidate suspension time bound for $\TASK{i}$. 
	\item \emph{Selector variable for candidate  suspension time:} For each $\TASK{i} \in \TASKSET$, for each segment $\REGION{i}{j} \in \ACCREGIONSET{i}$, for each $\nu_{i,g} \in \SCHEDPOINTS{i}$, $\ENSUSPVAR{i,j,g}  \in \{0,1\}$ is binary variable set to 1 if $\nu_{i,g}$ is the candidate suspension time bound chosen by the solver. 
\end{itemize}

\subsubsection{Constraints}

As discussed in Section~\ref{sec:npfpanalysis}, the suspension time can be bounded by leveraging Lemma~\ref{lemma:HWA_FP} and Equation~\eqref{eq:npfp_suff}.
Equation~\eqref{eq:npfp_suff} requires bounding the blocking from lower-priority tasks and the interference from high-priority tasks.   
Constraint~\ref{constraint:npblocking} bounds the former, enforcing the definition of variable $\BLOCKVAR{i}$. 

\begin{constraint} [NP blocking time] 
	\label{constraint:npblocking} 
	
	For each task $\TASK{i} \in \TASKSET$ such that $\REGIONSET{i}^H \neq \emptyset$, for each task $\TASK{s} \in \TASKSET\setminus{\TASK{i}}$, for each segment $\REGION{s}{f} \in \REGIONSET{s}^H$, 
	$$
	\BLOCKVAR{i} \ge    \EXECR{s}{f}{\ACCELERATOR} - (2 - \ACV{s,f} - \HIGHERPRIO{i}{s}) \cdot \BIGM. $$

\end{constraint}
\begin{proof} 
	The variables $\BLOCKVAR{i}$ bound the blocking time due to lower-priority tasks that any segment of $\TASK{i}$ can experience when accelerated on $\ACCELERATOR$ due to non-preemptive scheduling. 
	Following Lemma~\ref{lemma:HWA_FP}, such blocking is bounded by 	
	$$B_{i} = \max\left\{\EXECR{l}{v}{\ACCELERATOR}\, | \, \TASK{l} \in \LPSET{\TASK{i}}{} \wedge \REGION{l}{v} \in  \REGIONSET{l}^A\right\}.$$ 
	The constraint is enforced by requiring $\BLOCKVAR{i}$ to be greater than or equal to the WCET (on the accelerator) of all the segments that \textbf{(i)} are accelerated (i.e., $\ACV{s,f} = 1$), and \textbf{(ii)} are parts of a task $\TASK{s}$ with a lower priority than $\TASK{i}$ (i.e., $\HIGHERPRIO{i}{s} = 1$). In all the other cases, the constraint is disabled ($\BLOCKVAR{i} \ge -\infty$).
\end{proof}  

Constraint~\ref{constraint:interferingtaskinaccelerator} enforces the definition of the variables $\EACCVAR{i}{s}$, which represent the WCET of $\TASK{i}$ on $\ACCELERATOR$, if $\TASK{i}$ can interfere with $\TASK{s}$. 
\begin{constraint} [WCET of an interfering task on $\ACCELERATOR$] 
	\label{constraint:interferingtaskinaccelerator} 	
	For \\ each task $\TASK{i} \in \TASKSET$, for each segment $\REGION{i}{j} \in \REGIONSET{i}^H$, for each task $\TASK{s}  \in \TASKSET \setminus \TASK{i}$ such that $\REGIONSET{s}^H \neq \emptyset$,
	$$
	\EACCRVAR{i}{j}{s} \ge \EXECR{i}{j}{\ACCELERATOR} - (2 - \ACV{i,j} - \HIGHERPRIO{i}{s}) \cdot \BIGM, $$
	and for each $\TASK{i} \in \TASKSET$, $\TASK{s}  \in \TASKSET \setminus \TASK{i}$,
	$$\EACCVAR{i}{s} \ge \sum_{\REGION{i}{j} \in \ACCREGIONSET{i}} \EACCRVAR{i}{j}{s}$$
\end{constraint}
\begin{proof} 
 	The interfering WCET due to $\TASK{i}$ on a lower priority task $\TASK{s}$ is the sum of the individual contributions due to all segments $\REGION{i}{j} \in \REGIONSET{i}$.
 	Each segment contributes to the interference if: \textbf{(i)} $\REGION{i}{j}$ is accelerated ($\ACV{i,j} = 1$), and \textbf{(ii)} $\TASK{i}$ has higher priority than $\TASK{s}$ ($\HIGHERPRIO{i}{s} = 1$). The constraint follows.
\end{proof}

Constraint~\ref{constraint:suspensionbound} bounds the overall suspension time using Equation~\eqref{eq:npfp_suff}. 
This constraint follows alike to Constraint~\ref{constraint:WCRTbound}, but, in this case, we consider only those tasks that have at least one segment that can be accelerated as potential interfering tasks. 
The jitter component of Equation~\eqref{eq:npfp_suff} depends on the WCET of the task on the accelerator where it runs.  
Since the response-time bound is monotonic non-decreasing with respect to the jitter bound, and it is monotonic non-increasing with the WCET of the interfering task, we consider the minimum WCET $\WCETMINHW{i}$ with which a task $\TASK{i}$ can be characterized on the accelerator when $\ACCREGIONSET{i} \neq \emptyset$, when at least one of the segments in $\ACCREGIONSET{i}$ is accelerated. It is defined as
$$ \WCETMINHW{i}= \left\{\begin{array}{cl}
\min_{\REGION{i}{j} \in \ACCREGIONSET{i}} \EXECR{i}{j}{\ACCELERATOR} & \mbox{if } \REGIONSET{i}^W = \emptyset \\
\sum_{\REGION{i}{j} \in \REGIONSET{i}^W} \EXECR{i}{j}{\ACCELERATOR} & \mbox{otherwise, }
\end{array}
\right.
$$  
where $\REGIONSET{i}^W\subseteq \ACCREGIONSET{i}$ is the set of all segments of task $\TASK{i}$ that are \emph{necessarily} accelerated because they have only an implementation on $\ACCELERATOR$ (i.e., $\CANACCEL{i,j} = \mbox{\texttt{HWA}}$). 
This minimum WCET is then used to build the initial release jitters for such tasks $\TASK{s}$ with $\TASK{s}\in\TASKSET\setminus{\TASK{i}}$, such that $\REGIONSET{s}^H\neq \emptyset$, as follows:
\begin{equation}
J_s = \DEADLINE{s} - \WCETMINHW{s}.
\end{equation}
The corresponding checkpoint $V_{i,s}$ can then be computed again as in Theorem~\ref{th:lastactinst}.
For notation purposes we define the set of jitters at the hardware accelerator, for the tasks that can be accelerated, as
\begin{equation}\label{eq:jitterACC}
\overline{\mathcal{J}}^A_i =  \bigcup_{{\TASK{s}\in\TASKSET\setminus{\TASK{i}}\,|\, \REGIONSET{s}^H\neq \emptyset}} \left\{  \DEADLINE{s} - \WCETMINHW{s}\right\},
\end{equation} 
and the corresponding set of checkpoints as $ \SCHEDPOINTS{i}(\overline{\mathcal{J}}^A_i)$.

\begin{constraint} [Suspension bound candidate] \label{constraint:suspensionbound}
	For each\\ task $\TASK{i} \in \TASKSET$, for each segment $\REGION{i}{j} \in \ACCREGIONSET{i}$, for each $\nu_{i,g} \in \SCHEDPOINTS{i}(\overline{\mathcal{J}}^A_i)$, (see Eq.~\eqref{eq:npfp_suff})
	\begin{align*}
	&\STCVAR{i,j,g} \ge \BLOCKVAR{i} +  \sum_{\substack{\TASK{s}\in\TASKSET\setminus{\TASK{i}}\\ \REGIONSET{s}^H\neq \emptyset}}\ceilfrac{\nu_{i,g} +  J_{s}}{\PERIOD{s}}\cdot \EACCVAR{s}{i}, \\
	&\STCVAR{i,j,g} \le \nu_{i,g} + (1 - \ENSUSPVAR{i,j,g}) \cdot \BIGM,\\
	&\SR{i,j} \ge \EXECR{i}{j}{\ACCELERATOR} + \STCVAR{i,j,g} - (1 - \ENSUSPVAR{i,j,g}) \cdot \BIGM,
	\end{align*}
	For each task $\TASK{i} \in \TASKSET$, for each segment $\REGION{i}{j} \in \ACCREGIONSET{i}$ 	
	$$\sum_{\nu_{i,g} \in \SCHEDPOINTS{i}(\overline{\mathcal{J}}^A_i)} \ENSUSPVAR{i,j,g} = \ACV{i,j}.$$   
\end{constraint}
\begin{proof} 
	This constraint follows similarly to Constraint~\ref{constraint:WCRTbound}. 
	For each $\TASK{i} \in \TASKSET$ and for each segment $\REGION{i}{j} \in \ACCREGIONSET{i}$, there are up to  $|\SCHEDPOINTS{i}|$ candidate suspension-time bounds, represented with variables $\STCVAR{i,j,g}$ with $\nu_{i,g} \in \SCHEDPOINTS{i}(\overline{\mathcal{J}}^A_i)$.
	The fourth equality enforces only one of them to be selected as the actual suspension-time bound if the segment is accelerated.
	For each $\nu_{i,g} \in \SCHEDPOINTS{i}(\overline{\mathcal{J}}^A_i)$, the first inequality encodes Equation~\eqref{eq:npfp_suff} 
	bounding the jitter as $\JITs{s}{} = \DEADLINE{s} - \WCETMINHW{s}$.
	The second inequality enforces that the selected suspension-time bound must be valid according to Equation~\eqref{eq:npfp_suff} (see Section~\ref{sec:npfpanalysis}). In all other cases, the inequality is disabled (i.e., $\STCVAR{i,j,g} \le \infty$). 
	The third inequality enforces that the suspension-time bound is greater than or equal to the selected suspension-time candidate: otherwise, it is disabled (i.e., $\STCVAR{i,j,g} \ge -\infty$). 
\end{proof} 

Finally, Constraint~\ref{constraint:accelerationtimetask1}  must be enforced to bound the overall time spent on the hardware accelerator by each task $\TASK{i} \in \TASKSET$ and therefore the overall suspension time.
 
 \begin{constraint}[Suspension time of a task]
	\label{constraint:accelerationtimetask1} 
	For each task $\TASK{i} \in \TASKSET$, such that $\ACCREGIONSET{i} \neq \emptyset$,  
	$$ \ST{i} \ge \sum_{\REGION{i}{j} \in \ACCREGIONSET{i}} \SR{i,j}. $$
	Conversely, for each task  $\TASK{i} \in \TASKSET$, such that $\ACCREGIONSET{i} = \emptyset$,
	$$ \ST{i} = 0 $$
\end{constraint}

\subsection{Objective Function}
\label{sec:objectivefunction}

We design our optimization problem with four possible (alternative) objective functions. Two of them are devised to optimize the end-to-end latencies of individual chains, while the others consider the ratio between WCRT bounds and deadlines of individual tasks. 

For those targeting end-to-end latencies, we introduce, for each $\CHAIN{x} \in \CHAINSET$, the variables $\LAT{x} \in\mathbb{R}^{\geq 0}$ to encode the latency of chain $\CHAIN{x}$, in accordance with Equation~\eqref{eqn:chainlate}.\footnote{Enforcing $\EELATENCY{x} \ge \sum_{\TASK{i} \in \CHAIN{x}} (\R{i}{} + \PERIOD{i}) - \PERIOD{\mathit{first}}$ if sufficient because all objective functions are minimizations.}

\paragraph*{Min-Max Latency}

The first objective function encodes the goal of minimizing the maximum latency of all chains $\CHAIN{x} \in \CHAINSET$. It is defined as:
\begin{equation}\label{eq:lmax}
\mathbf{minimize}~~\MAXLAT,
\end{equation} 

where $\MAXLAT\in\mathbb{R}^{\geq 0}$ is a real variable that encodes the maximum latency of all chains $\CHAIN{x} \in \CHAINSET$ by enforcing the constraint $\MAXLAT \ge \LAT{x}$, $\forall \CHAIN{x} \in \CHAINSET$. 

\paragraph*{Min-Sum Latency}

The second objective function encodes the goal of minimizing the sum of the latency due to all the chains $\CHAIN{x} \in \CHAINSET$. It is defined as:
\begin{equation}\label{eq:lsum}
\mathbf{minimize} \sum_{\CHAIN{x} \in \CHAINSET} \LAT{x}
\end{equation}

\paragraph*{Min-Max WCRT-ratio}

The third objective function encodes the goal of minimizing the maximum ratio between the WCRT bound and the deadline of all tasks $\TASK{i} \in \TASKSET$. It is defined as:
\begin{equation}\label{eq:wcrtmax}
\mathbf{minimize}~~\RTMAX,
\end{equation} 

where $\RTMAX \in \mathbb{R}^{\geq 0}$ is a real variable that encodes the maximum ratios between the WCRT bound and the deadline of all tasks $\TASK{i} \in \TASKSET$ by enforcing the constraint $\RTMAX \ge \frac{\RTVAR{i}}{\DEADLINE{i}}$, $\forall \TASK{i} \in \TASKSET$.

\paragraph*{Min-Sum WCRT-ratio}

The last objective function encodes the goal of minimizing the sum of the ratios between the WCRT bound and the deadline of all tasks $\TASK{i} \in \TASKSET$. It is defined as:
\begin{equation}\label{eq:wcrtsum}
\mathbf{minimize} \sum_{\TASK{i} \in \TASKSET}  \frac{\RTVAR{i}}{\DEADLINE{i}}. 
\end{equation}

\section{Evaluation}
\label{sec:evaluation}

\begin{table*}
	\centering
	\caption{Parameters and solutions of the optimization problem for the task set provided with the challenge model. All times are in milliseconds.}
	\label{table:tasks}
	\begin{tabular}{|c|c|c|c|c|c|c|c||c|c|c|} 
		\hline
		\multicolumn{3}{|c|}{}                  & \multicolumn{2}{c|}{A57}       & \multicolumn{2}{c|}{Denver}    & \multicolumn{1}{c||}{GPU} & \multicolumn{3}{c|}{Min-Max Lat. RR Sol.}  \\ 
		\hline
		ID & Name           & $\PERIOD{i}$ (ms) & $\WCETNACC{i}$ & $\WCETACC{i}$ & $\WCETNACC{i}$ & $\WCETACC{i}$ & $C_{i}$                  & PRIO & CPU & ACC                           \\ \hline
		\hline
		1  & Lidar Grabber  & 33                & 14,379         & -             & 10,868         & -             & -                        & 8    & 5   & NO                            \\ 
		\hline
		2  & DASM           & 5                 & 1,958          & -             & 1,3            & -             & -                        & 5    & 1   & NO                            \\ 
		\hline
		3  & CAN Polling    & 10                & 0,632          & -             & 0,6            & -             & -                        & 1    & 1   & NO                            \\ 
		\hline
		4  & EKF            & 15                & 5,011          & -             & 4,430          & -             & -                        & 3    & 0   & NO                            \\ 
		\hline
		5  & Planner        & 12                & 13,939         & -             & 12,437         & -             & -                        & 2    & 2   & NO                            \\ 
		\hline
		6  & SFM            & 33                & 31,055         & 8,320         & 27,812         & 6,711         & 7,900                    & 4    & 3   & NO                            \\ 
		\hline
		7  & Localization   & 400               & 407,811        & 18,568        & 294,808        & 14,516        & 124,000                  & 7    & 4   & NO                            \\ 
		\hline
		8  & Lane Detection & 66                & 53,732         & 8,667         & 42,238         & 7,626         & 27,333                   & 6    & 5   & NO                            \\ 
		\hline
		9  & Detection      & 200               & -              & 4,958         & -              & 4,086         & 116,000                  & 0    & 0   & YES                           \\
		\hline
	\end{tabular}
\end{table*}

\begin{table*}
	\centering
	\caption{Running times.}
	\label{table:time}
	\begin{tabular}{|c|c|c|c|c|c|c|c|c|} 
		\hline
		& \multicolumn{2}{c|}{\textbf{Min-Max Lat}} & \multicolumn{2}{c|}{\textbf{Min-Sum Lat}} & \multicolumn{2}{c|}{\textbf{Min-Max RT}} & \multicolumn{2}{c|}{\textbf{Min-Sum RT}}  \\ 
		\hline
		& \textbf{Time} & \textbf{Opt.}             & \textbf{Time} & \textbf{Opt.}             & \textbf{Time} & \textbf{Opt.}            & \textbf{Time} & \textbf{Opt.}             \\ 
		\hline
		\textbf{No Contention} & 4.13s         & YES                       & 1h            & 2.41\%                    & 0.94s         & YES                      & 1h            & 11.81\%                   \\ 
		\hline
		\textbf{Round-Robin}   & 13.05s        & YES                       & 1h            & 0.18\%                    & 2.11s         & YES                      & 1h            & 1.68\%                    \\ 
		\hline
		\textbf{NP-FP}         & 2.98s         & YES                       & 26.2m         & YES                       & 1.45s         & YES                      & 1h            & 0.02\%                    \\
		\hline
	\end{tabular}
\end{table*}

We evaluate the proposed optimization method on the WATERS 2019 Challenge by Bosch~\cite{WATERS2019}.
As discussed in Section~\ref{sec:waters}, the WATERS 2019 Challenge Platform Model is based on the NVIDIA Jetson TX2. This heterogeneous platform is composed of a quad-core 1.9GHz ARMv8 A57, a dual-core 2GHz ARMv8 Denver, and an integrated GPU.

Consequently, the proposed platform model comprises two types of processor cores: the first four cores $\COREC{k} \in \PROCESSORSET$, with $k\in\{0,1,2,3\}$ are the \textit{A57} cores, while the last two, i.e., $\COREC{k} \in \PROCESSORSET$, with $k\in\{4,5\}$ are the \textit{Denver} cores.  

Table~\ref{table:tasks} reports the attributes of the WATERS 2019 task set. Each task includes a single segment. Tasks \texttt{SFM}, \texttt{Localiza-}
\texttt{tion}, and \texttt{Lane Detection} are provided with both a fully CPU-based implementation and a GPU-based implementation, while \texttt{Detection} is provided only with an accelerated implementation. All others tasks need to run on CPU cores.
For each type of processor, Table~\ref{table:tasks} reports $\WCETNACC{i}$ and $\WCETACC{i}$ to denote the overall WCET of the task when it is running solely on a core or using the accelerator, respectively. Similarly, the column labeled with $C_i$ under the GPU group reports the overall tasks' WCET when executing on the GPU, if the task is accelerated. 
The period (equal to the relative deadline) is also reported. 
All times are in milliseconds. 

In such a scenario, our MILP formulation determines: \textbf{(i)} which tasks to accelerate on the GPU, \textbf{(ii)} which tasks to execute on the (faster) Denver cores, and which to execute on A57 cores,  \textbf{(iii)} the task-to-core assignment, \textbf{(iv)} the priority assignment. The corresponding solution needs to guarantee the application timing constraints, i.e., each task needs to complete within its deadline.

The proposed MILP formulation has been coded in C++ and solved with IBM CPLEX on a machine equipped with an  Intel Core i7-6700K @ 4.00GHz.

We compared the four different objective functions discussed in Section~\ref{sec:objectivefunction}.  
For each of them, we considered three different policy for the hardware accelerator: \textbf{(i)} round-robin, \textbf{(ii)} non-preemptive fixed-priority and \textbf{(iii)} no contention, where each accelerated task runs in the accelerator without interference.  
The third case serves both for baseline comparison and it can give some intuitions on the results that can be obtained by the optimization problem for hardware accelerators that do not provide time interference (e.g., dedicated, statically-deployed FPGA-based accelerators where interference-related delays are only due to contention for memories, interconnects, and devices\footnote{Note that performing an optimization for FPGA-based accelerators requires additional considerations about the physical resource requirements on the programmable logic~\cite{Seyoum2019}.}).

Figure~\ref{fig:exps_rt} reports the ratio between: \textbf{(i)} the obtained WCRT and the deadline (RT/D) and, \textbf{(ii)} the suspension-time bound and the deadline (ST/D) for such configurations.
For the configuration of Figure~\ref{fig:exps_rt}(a) the last three columns of Table~\ref{table:tasks} show the priority assigned by the solver, the task-to-core assignment, and whether the task has been selected for being accelerated.
For all the objective functions, the solver accelerates only the \texttt{Detection} task both for round-robin and non-preemptive fixed priorities as it deems more convenient to run the other three accelerable tasks on a CPU core rather than congesting the GPU, while in the ``no contention'' case all tasks are accelerated due to the shorter WCETs they experience by running on the GPU.  By comparing inset (a) with (b), and (c) with (d), we note that objective functions minimizing the maximum RT/D and chain latency tend to provide higher ratios, as in this case, the solver needs to optimize only the task leading to the maximum RT/D ratio, while objective functions minimizing the sum of RT/D and chain latencies generally provide lower values of the RT/D ratio. For example, this is the case of the \texttt{CAN} task, which achieves an RT/D of 1 in insets (a) and (c), and a very small RT/D in insets (b) and (d). 
On the other hand, the Min-Sum objectives are more difficult problems to solve: indeed, they manifested also higher running times. Table~\ref{table:time} shows the running times achieved by running CPLEX with a timeout of 1 hour. For the Min-Sum objectives, the solver found the optimal solution within the allowed time in only one case. However, in almost all cases, the solver provides a solution very close to optimality, with an optimality gap below 3\%. Only for the Min-Sum RT objective in the ``no contention'' case, the optimality gap after 1-hour running is 11.81\%: this is attributed to the fact that in the no contention case, there are fewer constraints to impose, and thus a larger search space. 

Figure~\ref{fig:exps_rt_scaled} reports another interesting configuration we found in our experimentation. In this case, the WATERS 2019 WCETs have been scaled to 80\% of their values: this leads the optimizer to accelerate also the \texttt{Localization} task with the round-robin policy. Indeed, with smaller WCETs, the interference imposed by \texttt{Localization} on \texttt{Detection} becomes smaller, making it convenient to accelerate also \texttt{Localization}, which has a way smaller WCET when running on the accelerator (see Table~\ref{table:tasks}). Therefore, the proposed optimization problem allows recognizing non-trivial trade-offs, allowing to obtain optimal solutions that would be very hard to grasp without relying on real-time analysis.

Finally, Table~\ref{table:chains} reports the chain latencies obtained with the four different objective functions. The second column of the table reports an ordered list of task IDs (introduced in Table~\ref{table:tasks}) representing the tasks in the chain.
As expected, the longest chains are those involving the \texttt{Localization} task, which has a large period (see Equation~\eqref{eqn:chainlate}).
From the results, we can observe that chain-specific objective functions may lead to greatly smaller latency values: for example, this is the case of chain C4, which achieves a latency of about $760$ ms with the Min-Max Lat objective and much higher latency of about $848$ ms with the Min-Max RT.

\newcommand{\PlotHeight}{80}
\newcommand{\PlotWidth}{2}

\pgfplotsset{compat=1.13}
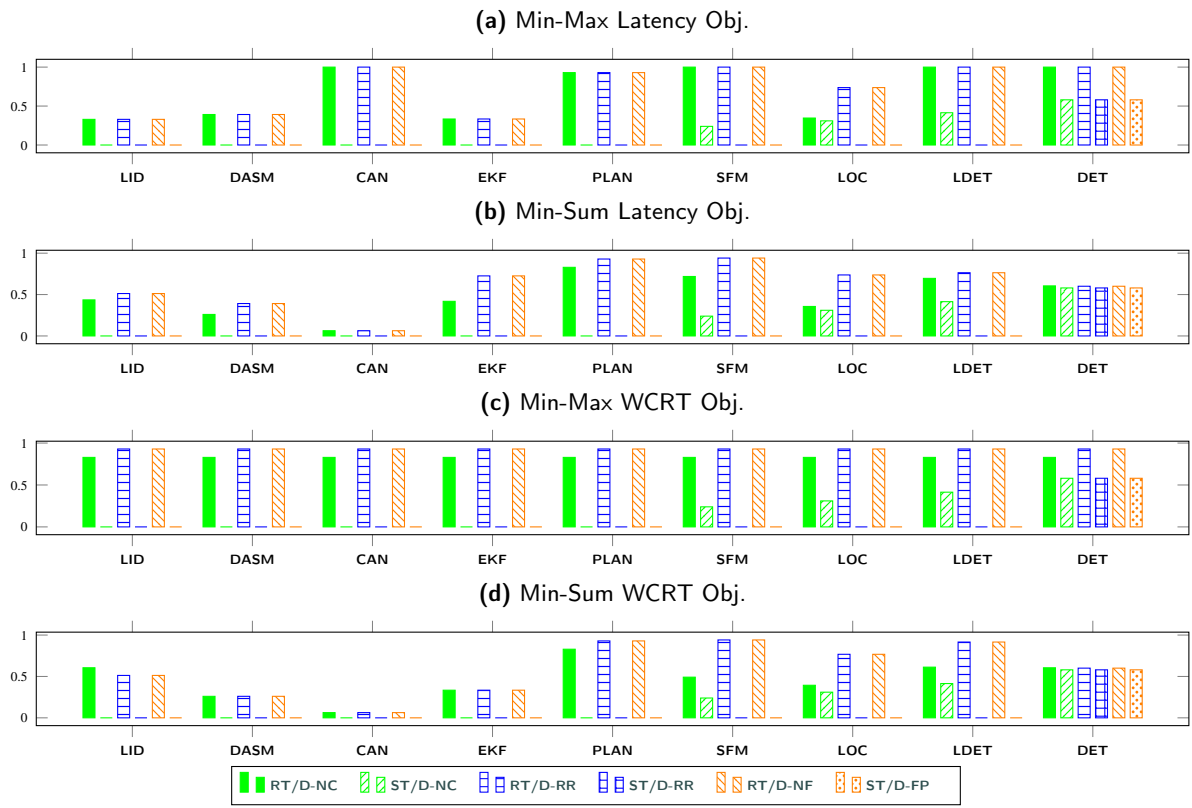
\begin{figure*}[pos=t,width=\textwidth,align=\centering] 
		\begin{center}
			
\pgfplotstableread[col sep=comma]{plots/rt_chainmax.txt}\mydata
\begin{tikzpicture}
\begin{axis}[width=\PlotWidth\columnwidth, height=\PlotHeight, align =center,  
name={theaxis},
title = {\textbf{(a)} Min-Max Latency Obj.},
xticklabels={ \textbf{\tiny LID}, \textbf{DASM}, \textbf{CAN}, \textbf{EKF}, \textbf{PLAN}, \textbf{SFM}, \textbf{LOC}, \textbf{LDET}, \textbf{DET}},xtick={0,...,8},
ticklabel style = {font=\tiny},
ybar,
bar width=0.1,
legend columns=-1,
legend entries={{\tiny \textbf{RT/D-NC}\quad\quad}, {\tiny \textbf{ST/D-NC}\quad\quad}, {\tiny \textbf{RT/D-RR}\quad\quad}, {\tiny \textbf{ST/D-RR}\quad\quad},{\tiny \textbf{RT/D-NF}\quad\quad}, {\tiny \textbf{ST/D-FP}\quad\quad}},  
legend to name=leg]  
]
\addplot [color=green, fill] table [x expr=\coordindex, y={RTRN}] \mydata;
\addplot [color=green, pattern=north east lines, pattern color=green] table [x expr=\coordindex, y={STRN}] \mydata; 
\addplot [color=blue, pattern=horizontal lines, pattern color=blue] table [x expr=\coordindex, y={RTRR}] \mydata;
\addplot [color=blue, pattern=grid, pattern color=blue] table [x expr=\coordindex, y={STRR}] \mydata; 
\addplot [color=orange, pattern=north west lines, pattern color=orange] table [x expr=\coordindex, y={RTRF}] \mydata;
\addplot [color=orange, pattern=crosshatch dots, pattern color=orange] table [x expr=\coordindex, y={STRF}] \mydata; 
\end{axis}
\end{tikzpicture}  
\pgfplotstableread[col sep=comma]{plots/rt_chainsum.txt}\mydata
\begin{tikzpicture}
\begin{axis}[width=\PlotWidth\columnwidth, height=\PlotHeight, align =center,  
name={theaxis},
title = {\textbf{(b)} Min-Sum Latency Obj.},
xticklabels={ \textbf{\tiny LID}, \textbf{DASM}, \textbf{CAN}, \textbf{EKF}, \textbf{PLAN}, \textbf{SFM}, \textbf{LOC}, \textbf{LDET}, \textbf{DET}},xtick={0,...,8},
ticklabel style = {font=\tiny},
ybar,
bar width=0.1,
]
\addplot [color=green, fill] table [x expr=\coordindex, y={RTRN}] \mydata;
\addplot [color=green, pattern=north east lines, pattern color=green] table [x expr=\coordindex, y={STRN}] \mydata; 
\addplot [color=blue, pattern=horizontal lines, pattern color=blue] table [x expr=\coordindex, y={RTRR}] \mydata;
\addplot [color=blue, pattern=grid, pattern color=blue] table [x expr=\coordindex, y={STRR}] \mydata; 
\addplot [color=orange, pattern=north west lines, pattern color=orange] table [x expr=\coordindex, y={RTRF}] \mydata;
\addplot [color=orange, pattern=crosshatch dots, pattern color=orange] table [x expr=\coordindex, y={STRF}] \mydata; 
\end{axis}
\end{tikzpicture}
\pgfplotstableread[col sep=comma]{plots/rt_rtmax.txt}\mydata
\begin{tikzpicture}
\begin{axis}[width=\PlotWidth\columnwidth, height=\PlotHeight, align =center,  
name={theaxis},
title = {\textbf{(c)} Min-Max WCRT Obj.},
xticklabels={ \textbf{\tiny LID}, \textbf{DASM}, \textbf{CAN}, \textbf{EKF}, \textbf{PLAN}, \textbf{SFM}, \textbf{LOC}, \textbf{LDET}, \textbf{DET}},xtick={0,...,8},
ticklabel style = {font=\tiny},
ybar,
bar width=0.1,
]
\addplot [color=green, fill] table [x expr=\coordindex, y={RTRN}] \mydata;
\addplot [color=green, pattern=north east lines, pattern color=green] table [x expr=\coordindex, y={STRN}] \mydata; 
\addplot [color=blue, pattern=horizontal lines, pattern color=blue] table [x expr=\coordindex, y={RTRR}] \mydata;
\addplot [color=blue, pattern=grid, pattern color=blue] table [x expr=\coordindex, y={STRR}] \mydata; 
\addplot [color=orange, pattern=north west lines, pattern color=orange] table [x expr=\coordindex, y={RTRF}] \mydata;
\addplot [color=orange, pattern=crosshatch dots, pattern color=orange] table [x expr=\coordindex, y={STRF}] \mydata; 
\end{axis}
\end{tikzpicture}

\pgfplotstableread[col sep=comma]{plots/rt_rtsum.txt}\mydata
\begin{tikzpicture}
\begin{axis}[width=\PlotWidth\columnwidth, height=\PlotHeight, align =center,  
name={theaxis},
title = {\textbf{(d)} Min-Sum WCRT Obj.},
xticklabels={ \textbf{\tiny LID}, \textbf{DASM}, \textbf{CAN}, \textbf{EKF}, \textbf{PLAN}, \textbf{SFM}, \textbf{LOC}, \textbf{LDET}, \textbf{DET}},xtick={0,...,8},
ticklabel style = {font=\tiny},
ybar,
bar width=0.1,
]
\addplot [color=green, fill] table [x expr=\coordindex, y={RTRN}] \mydata;
\addplot [color=green, pattern=north east lines, pattern color=green] table [x expr=\coordindex, y={STRN}] \mydata; 
\addplot [color=blue, pattern=horizontal lines, pattern color=blue] table [x expr=\coordindex, y={RTRR}] \mydata;
\addplot [color=blue, pattern=grid, pattern color=blue] table [x expr=\coordindex, y={STRR}] \mydata; 
\addplot [color=orange, pattern=north west lines, pattern color=orange] table [x expr=\coordindex, y={RTRF}] \mydata;
\addplot [color=orange, pattern=crosshatch dots, pattern color=orange] table [x expr=\coordindex, y={STRF}] \mydata; 
\end{axis}
\end{tikzpicture}
\ref{leg}
\end{center}

\caption{Ratio between the obtained WCRT bounds and the deadline, and the suspension time bounds and the deadline for each task of the WATERS 2019 Challenge obtained by running the MILP with different objective functions reported above each graph.
}
\label{fig:exps_rt}

\end{figure*}

\renewcommand{\PlotHeight}{80}
\renewcommand{\PlotWidth}{2}
{
\pgfplotsset{compat=1.13}

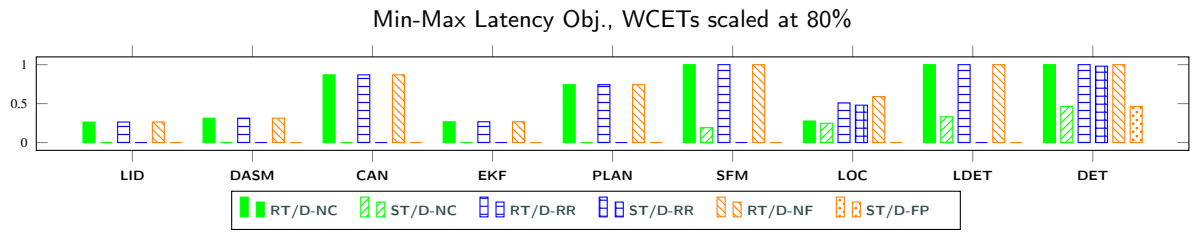
\begin{figure*}[pos=t,width=\textwidth,align=\centering] 
	\begin{center}
		\pgfplotstableread[col sep=comma]{plots/80/rt_chainmax_80.txt}\mydata
		\begin{tikzpicture}
		\begin{axis}[width=\PlotWidth\columnwidth, height=\PlotHeight, align =center,  
		name={theaxis},
		title = {Min-Max Latency Obj., WCETs scaled at 80\%},
		xticklabels={ \textbf{\tiny LID}, \textbf{DASM}, \textbf{CAN}, \textbf{EKF}, \textbf{PLAN}, \textbf{SFM}, \textbf{LOC}, \textbf{LDET}, \textbf{DET}},xtick={0,...,8},
		ticklabel style = {font=\tiny}, 
		ybar,
		bar width=0.1,
		]
		\addplot [color=green, fill] table [x expr=\coordindex, y={RTRN}] \mydata;
		\addplot [color=green, pattern=north east lines, pattern color=green] table [x expr=\coordindex, y={STRN}] \mydata; 
		\addplot [color=blue, pattern=horizontal lines, pattern color=blue] table [x expr=\coordindex, y={RTRR}] \mydata;
		\addplot [color=blue, pattern=grid, pattern color=blue] table [x expr=\coordindex, y={STRR}] \mydata; 
		\addplot [color=orange, pattern=north west lines, pattern color=orange] table [x expr=\coordindex, y={RTRF}] \mydata;
		\addplot [color=orange, pattern=crosshatch dots, pattern color=orange] table [x expr=\coordindex, y={STRF}] \mydata; 
		\end{axis}
		\end{tikzpicture}
		\ref{leg}
			\end{center} 
		\caption{Ratio between the obtained WCRT bounds and the deadline, and the suspension time bounds and the deadline for each task of the WATERS 2019 Challenge obtained by running the MILP with the Min-Max Latency objective function and WCET scaled at the 80\% of their values.}
		\label{fig:exps_rt_scaled}
\end{figure*}
}

\begin{table*}
	\centering
	\caption{Latencies (in ms) of the processing chains with the hardware accelerator adopting the NP-FP policy.}
	\label{table:chains}
	\begin{tabular}{|c|c|c|c|c|c|} 
		\hline
		ID & Tasks             & Min-Max Lat & Min-Sum Lat & Min-Max RT        & Min-Sum RT  \\ 
		\hline
		C1 & 9 - 5 - 2         & 235,897    & 155,983    & 224,438     & 155,325    \\ 
		\hline
		C2 & 6 - 5 - 2         & 68,897     & 66,952     & 69,251      & 66,294    \\ 
		\hline
		C3 & 8 - 5 - 2         & 101,897    & 86,307     & 99,916 & 95,677     \\ 
		\hline
		C4 & 3 - 7 - 4 - 5 - 2 & 760,716    & 757,222    & 848,523   & 762,948    \\ 
		\hline
		C5 & 1 - 7 - 4 - 5 - 2 & 761,584    & 773,497    & 869,896          & 779,223    \\ 
		\hline
		C6 & 1 - 5 - 2         & 46,76     & 52,804   & 69,251      & 52,146     \\ 
		\hline
		C7 & 3 - 4 - 5 - 2     & 65,908     & 62,414     & 76,817           & 55,882     \\ 
		\hline
		C8 & 3 - 5 - 2         & 45,897     & 36,529     & 47,878           & 35,871     \\
		\hline
	\end{tabular}
\end{table*}
\section{Related Work}
\label{sec:related}

Heterogeneous platforms have received great attention in the last years, especially in the field of \emph{high-performance parallel computing}. The primary focus of researchers in this field has been dedicated to improving performance (e.g., by pursuing faster computation) and reducing energy consumption, mainly regarding mainstream computing, but with little attention to real-time constraints or embedded systems. A wide survey on this topic of partitioning techniques and benchmarks regarding combined CPU/GPU architectures can be found in~\cite{mittal2015survey}. Among the many, it is worth mentioning the work of Li et al.~\cite{li2021efficient}, which presents a set of algorithms for task mapping in heterogeneous platforms with GPUs, with the goal of minimizing the makespan.

The introduction of heterogeneous platforms and hardware accelerators is more recent for what concerns embedded real-time applications.
While most of the available works in real-time literature involving multicore platforms are often constrained to \emph{homogeneous} processors~\cite{DavisSurvey}, there is a significant portion of literature that started exploring the potential and challenges of heterogeneous platforms applied to the case of embedded real-time applications.

Concerning the study of specific accelerators in real-time systems, prior works mostly targeted GPU-based and FPGA-based acceleration.
Several efforts have been spent on improving the predictability of workloads running on GPU accelerators. Since the scheduling policies implemented in such accelerators are typically not disclosed by hardware vendors, a branch of research investigates their internal behavior~\cite{Amert2021,Amert2017,Sanudo2020}, e.g., through reverse engineering. Capodieci et al.~\cite{Capodieci2017}, Ali and Yun~\cite{Ali2018}, and Forsberg et al.~\cite{Forsberg2017} proposed mechanisms to achieve control on the memory traffic on platforms with GPUs.
Cavicchioli et al.~\cite{Cavicchioli2019} studied novel methods to offload computations to a GPU.
Other research provided an implementation of the constant bandwidth server on an NVIDIA GPU~\cite{Capodieci2018}. 

Concerning FPGA-based accelerators, Biondi et al.~\cite{Bio-RTSS16} introduced the FRED framework to support real-time applications on heterogeneous platforms. FRED provides a predictable infrastructure and a corresponding analysis to guarantee bounded delays when
requesting a dynamically reconfigured hardware accelerator. Later, Pagani et al.~\cite{Pagani2018} provided an implementation for FRED based on Linux. Other works aimed at improving the predictability of bus accesses of hardware-accelerated tasks on FPGA-based platforms and evaluating the profitability of performing acceleration requests~\cite{Pagani2019,Restuccia2020FCCM,Restuccia2020DAC,Restuccia2020,Valente2020, Wang2020}.

Considering one of the main problem addressed in this work, i.e., \emph{task partitioning} onto heterogeneous platforms,
in general, most of the available literature on the topic is limited to heterogeneous platforms composed of \emph{CPUs only}. 
The problem of partitioning real-time applications onto platforms with heterogeneous processing cores is known to be NP-hard in the strong sense~\cite{baruah2004partitioning}. This problem has been mainly addressed with the usage of integer linear programming (ILP) techniques~\cite{baruah2004partitioning, baruah2016ilp, baruah2019ilp}. Additionally, heuristics methods have been proposed to solve this problem, e.g., those based on an \emph{ant-colony optimization} approach~\cite{chen2011assigning} and an improved \emph{particle swarm optimization}~\cite{poongothai2014heuristic}. 
However, none of these methods considers the presence of hardware accelerators, which introduces considerable challenges, e.g., a suspension behavior on the processing cores and deciding whether it is convenient (for a WCRT perspective) to perform acceleration. 

In the context of platforms with heterogeneous cores, a popular target platform is the big.LITTLE from ARM~\cite{jeff2012big}, which is composed of a ``little'' and power-efficient set of cores, together with a ``big'' set of cores for high-performance computation. Partitioning heuristics exist for such a specific architecture, e.g., see~\cite{zahaf2017energy, mascitti2020heuristic, mascitti2020adaptive, mascitti2021}. 
In the works mentioned above, schedulability is checked by means of utilization-based tests.   

The problem of partitioning applications to multicore platforms also requires that the functional dependencies are preserved in the final deployment. Such dependencies are commonly coded in the form of precedence constraints between tasks, constituting a direct acyclic graph (DAG). The problem of partitioning DAGs on multicores while preserving their functional dependencies has been addressed, e.g., in~\cite{buttazzo2011partitioning}, and it also drew much attention recently in the automotive field~\cite{hottger2015model, lowinski2016splitting}. 
Time determinism and causality in multicore platforms can be effectively addressed with the usage of Logical Execution Time Paradigm~\cite{henzinger2003control}. Its practical application to multicore has been explored in~\cite{biondilogical, biondi2018achieving, Pazzaglia2021}, and specifically for addressing the problem of partitioning real-time applications on a (homogeneous) multicore in~\cite{pazzaglia2019optimizing}. The work in~\cite{guo2019energy} proposes a mapping of multiple DAG tasks that also minimizes power consumption, without considering hardware accelerators.
A response time analysis for a DAG task on a heterogeneous platform with a single hardware accelerator is presented in~\cite{serrano2018response}. Only applications performing a single acceleration request for the whole application (composed of a single DAG task) are supported.
Zahaf et al.~\cite{Zahaf2020} presented the HPC-DAG model and corresponding schedulability analysis, specifically considering NVIDIA platforms.

The introduction of shared devices in the platform, such as GPUs and FPGAs, brings additional complexity in the analysis  problem. Indeed, tasks accessing shared resources or performing operations on external devices are subject to \emph{suspension delays}, which must be properly accounted for to ensure schedulability~\cite{chen2019many}.  
Many analyses have been then devised on the topic, with different goals, e.g., supporting global scheduling techniques~\cite{Dong2016, liu2013suspension}, analyzing EDF scheduling~\cite{Gunzel2020}, providing a unifying analysis~\cite{Chen2016}, analyzing parallel tasks~\cite{Casini2018Time, Fonseca2017}, or supporting soft real-time tasks~\cite{Liu2010}. 
A comprehensive review of works about self-suspensions can be found in the two excellent surveys by Chen et~\cite{Chen2017,  chen2019many}.
A set of task partitioning algorithms in the presence of a shared resource, with the goal of guaranteeing schedulability while minimizing the required size of the shared resource, is presented in~\cite{dong2018shared}. No priority assignment is provided, and only homogeneous cores are considered.

Introducing self-suspensions also brings additional complexity in the formulation of the task mapping problem. For instance, the response time formulation must be properly adapted in order to be used in a linear optimization environment. In this work, we make use of an approximated analysis based on checking only a smart subset of time instants, following the work in~\cite{Pazzaglia2019DATE}. Other works addressed this problem with ad-hoc formulations for the response time.
An example of the ILP method that addresses optimal partitioning in the case of shared resources, together with a response time analysis (but on a \emph{homogeneous} multicore platform), is presented in~\cite{wieder2013efficient}. A heuristic for resource-oriented task partitioning in multicores is presented in~\cite{huang2016resource}. Other partitioning strategies are proposed in~\cite{chen2019improved}.
All these works, however, target only homogeneous multicores.

To the best of our knowledge, there is no prior work targeting the problem of providing an optimal solution for the partitioning and priority assignment of real-time tasks on heterogeneous platforms that also include an hardware accelerator.

\section{Conclusions}
\label{sec:conclusions}

This paper provides solutions for modeling, analyzing, and partitioning real-time applications running onto heterogeneous platforms equipped with a hardware accelerator.
We presented a task and platform model that is applicable to different use-cases, together with a response time analysis that supports different scheduling policies at the hardware-accelerator level, and which can be easily linearized to be used in an ad-hoc optimization problem.
The proposed approach is a simple and flexible way for finding an optimal task-to-core mapping and priority assignment for real-time autonomous applications, and for deciding whether to accelerate tasks provided with both CPU-based and accelerated implementations under different scheduling policies for the accelerator.
Nevertheless, there is plenty of space for future work in a flourishing research field as the one of heterogeneous real-time systems.
For example, interesting future research directions include the consideration of more precise analyses of self-suspending tasks~\cite{Nelissen2015}, the extension to asynchronous hardware acceleration~\cite{Aromolo2021}, ROS-based systems~\cite{Casini2019ECRTS} and frameworks for deep neural networks~\cite{Casini2020SPE,Casini2019DAC,Yaman2017,Pfenning2021}, and the study of specific hardware accelerators~\cite{Sanudo2020} and acceleration methods~\cite{Bio-RTSS16} to provide fine-grained bounds on the suspension time.

\section*{Acknowledgments}
This work has been partially supported by the EU H2020 project AMPERE under the grant agreement no. 871669.

\begin{spacing}{1}
\bibliographystyle{cas-model2-names}

	\bibliography{references}
\end{spacing}

\end{document}